\documentclass[10pt, aps, prd, nofootinbib, preprintnumbers,
  twocolumn, superscriptaddress]{revtex4-2}
\pdfoutput=1

\usepackage[pdftex]{graphicx}
\usepackage{amssymb, amsmath, bm}
\usepackage{microtype}
\usepackage{etoolbox}
\apptocmd{\sloppy}{\hbadness 10000\relax}{}{}

\usepackage{xcolor}

\usepackage[colorlinks,linktocpage,hypertexnames=false]{hyperref}
\hypersetup{colorlinks=true, citecolor=blue}

\begin{document}
\preprint{INR-TH-2025-011}
\title{Self-similar kinetics for gravitational Bose-Einstein condensation}

\author{A.S. Dmitriev}\email{dmitriev.as15@physics.msu.ru}
\affiliation{Institute for Nuclear Research of the Russian Academy
  of Sciences, Moscow 117312, Russia}
\author{D.G. Levkov}\email{levkov@ms2.inr.ac.ru}
\affiliation{Institute for Nuclear Research of the Russian Academy
  of Sciences, Moscow 117312, Russia}
\affiliation{Institute for Theoretical and Mathematical
  Physics, MSU, Moscow 119991, Russia}
\author{A.G. Panin}\email{panin@ms2.inr.ac.ru}
\affiliation{Institute for Nuclear Research of the Russian Academy
  of Sciences, Moscow 117312, Russia}
\author{I.I. Tkachev}\email{tkachev@ms2.inr.ac.ru}
\affiliation{Institute for Nuclear Research of the Russian Academy
  of Sciences, Moscow 117312, Russia}
  \affiliation{Novosibirsk State University, Novosibirsk 630090, Russia}

  \begin{abstract}
    We study an overpopulated gas of gravitationally interacting bosons
    surrounding a droplet of Bose-Einstein condensate~--- Bose star. We
    argue that kinetic evolution of this gas approaches with
    time a self-similar attractor solution to the kinetic
    equation. If the scale symmetry of the equation is broken by
    external conditions,  the attractor solution exists, remains
      approximately self-similar, but has slowly drifting scaling
    dimension. The latter new regime of {\it adiabatic
      self-similarity} can determine growth of dark matter Bose stars
    in cosmological models.
  \end{abstract}

  
\maketitle

\section{Introduction and summary}
\label{sec:intro}

\begin{figure}[t]
  \centerline{(a)}
  \centerline{\hspace{5mm}\includegraphics{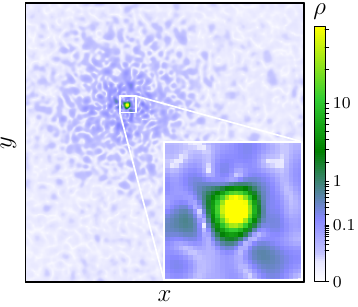}}

  \centerline{\unitlength=1mm
    \begin{picture}(80.7,40)
      \put(0,0){\includegraphics{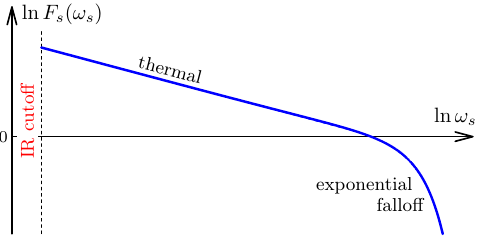}}
      \put(38.5,33){(b)}
  \end{picture}}
  \caption{(a) A minicluster from the simulation in
    Ref.~\cite{Dmitriev:2023ipv}. It includes a weakly bound
    virialized gas (clumpy spherical pattern) and a newborn Bose star
    (bright dot in the center). Color indicates dark matter 
    mass density~$\rho(x,\, y,\, z)$ at~${z=0}$ in dimensionless
    units$^{\mbox{\scriptsize \ref{foot-units}}}$, 
    the inset zooms onto the Bose star. (b)~Typical 
    self-similar profile~$F_s(\omega_s)$ in
    Eqs.~(\ref{eq:an}),~(\ref{eq:ab}) 
    (log-log scale).}
  \label{fig:schematic}
\end{figure}

Cosmological simulations reveal~\cite{Schive:2014dra, *Schive:2014hza,
  *Liao:2024zkj, Mocz:2017wlg, Levkov:2018kau, Veltmaat:2018dfz,
  *Schwabe:2021jne, Schwabe:2020eac, Eggemeier:2019jsu, Mina:2020eik,
  Amin:2025sla} that the smallest clumps of light bosonic
(axion-like)~\cite{RosenbergPDG} dark matter look unusual. Apart from
virialized dark gas they include Bose stars~\cite{Ruffini:1969qy,
  Tkachev:1986tr}~--- gravitationally self-bound drops of dark matter
Bose-Einstein condensate, see Fig.~\ref{fig:schematic}(a). If the dark
bosons are QCD axions~\cite{Sikivie:2006ni, Chadha-Day:2021szb}, this 
structure appears~\cite{Levkov:2018kau, Eggemeier:2019jsu} inside
asteroid-mass axion miniclusters~\cite{Kolb:1993zz, *Kolb:1993hw, 
  Vaquero:2018tib, Buschmann:2019icd, Ellis:2020gtq,
  Pierobon:2023ozb}, and if they are ultra-light
(fuzzy)~\cite{Marsh:2015xka, Ferreira:2020fam, *Eberhardt:2025caq, 
  Hui:2021tkt}, Bose stars emerge as gigantic solitonic
cores~\cite{Schive:2014dra, *Schive:2014hza, *Liao:2024zkj,
  Mocz:2017wlg, Veltmaat:2018dfz, *Schwabe:2021jne, Mina:2020eik,
  Amin:2025sla} of dwarf galaxies. In both scenarios Bose-Einstein
condensation is ensured~\cite{Sikivie:2009qn, *Erken:2011dz,
  Levkov:2018kau} by huge phase-space density
(overpopulation)~\cite{Preskill:1982cy, Abbott:1982af} of dark matter 
and its partial thermalization via long-range gravitational
forces. Indeed, birth  of Bose stars in simulations 
takes time comparable~\cite{Levkov:2018kau,  Chen:2020cef,
  *Chen:2021oot, Chan:2022bkz, Jain:2023tsr} to the gravitational relaxation
time~\cite{Levkov:2018kau}. The  question is, how do these objects 
grow further  by condensing dark bosons?  

\footnotetext{\label{foot-units}We measure~$\rho$
    and~$x,\, y$ in units of~$p_0^4/(m^2 G)$ and~$p_0^{-1}$,
    respectively, where~$m$ is the particle mass,~$p_0$ is the reference
    particle momentum, and~${\hbar=1}$; 
    see Sec.~\ref{sec:attr-self-simil} for details.} 
Recently, we observed~\cite{Dmitriev:2023ipv} that an overpopulated
dark gas  surrounding the Bose star evolves in a self-similar 
way. Namely, its distribution function~${F(t, \,\omega) = 
  dN/d\omega}$ of particles over energies~$\omega$
retains form during long evolution periods just getting rescaled
by coefficients~$\alpha(t)$ and~$\beta(t)$,
\begin{equation}  
  \label{eq:an}
  F(t,\, \omega)\equiv \frac{dN}{d\omega}= \alpha(t) F_s\left(\beta(t)
  \omega\right)\,.
\end{equation}
Moreover, the coefficients themselves are powers of time:
\begin{equation}
  \label{eq:ab}
  \alpha(t) \propto (t - t_i)^{-1/D}\quad \mbox{and} \quad \beta(t)
  \propto (t -
  t_i)^{2/D-1}\,,
\end{equation}
where~$D$ parametrizes scaling dimensions and~$t_i$ is the
time origin. The law~(\ref{eq:an}), (\ref{eq:ab}) was first discovered in 
microscopic (Schr\"odinger-Poisson) simulations and then derived from
simplified kinetic equation~\cite{Dmitriev:2023ipv}.

Scaling kinetic evolutions akin to Eqs.~(\ref{eq:an}), (\ref{eq:ab}) were
theoretically observed in a plethora of nonequilibrium systems
at large  occupancies: in Bose-condensing gases with short-range
interactions, both at weakly turbulent
(kinetic)~\cite{Svistunov:1991, *Svistunov:2001, 
  Falkovich:1991, Semikoz:1994zp, *Semikoz:1995rd, Berges:2012us, 
  Berges:2013lsa, SEMISALOV2021105903} and strongly coupled
(superfluid)~\cite{Kagan:1994, Berloff:2002, Nowak:2011sk,
  Nowak:2012gd,  Chantesana:2018qsb, Mikheev:2018adp, Schmied:2018mte}
stages of this process, in non-Abelian plasmas created by heavy ion
collisions~\cite{Kurkela:2012hp, Berges:2013eia, *Berges:2013fga,
  *Berges:2014bba, AbraaoYork:2014hbk, PineiroOrioli:2015cpb,
  Kurkela:2015qoa}, and in stochastic matter fields thermalizing
at the post-inflationary epoch~\cite{Micha:2002ey, *Micha:2004bv,
  Berges:2008wm}. Self-similar dynamics of ultra-cold gases is
already confirmed by table-top experiments~\cite{Prufer:2018hto,
  Erne:2018gmz, Glidden:2020qmu, GarciaOrozco:2021, Madeira:2022}. 

Physically, self-similar kinetic solutions can play the role of
``nonthermal attractors''~\cite{Berges:2008wm} visited by the system
on its way to thermal equilibrium. They may delay  thermalization to
many relaxation times~\cite{Micha:2002ey, *Micha:2004bv}. In addition,
these solutions can transport conserved quantities to low or
high momenta thus generalizing~\cite{Falkovich:1991,
  Semikoz:1994zp, *Semikoz:1995rd} stationary Kolmogorov-Zakharov  
cascades~\cite{Zakharov:1966, Nazarenko:2011, zakharov2012kolmogorov}.
Usually, theoretical studies of self-similar dynamics boil down to
  computation of scaling exponents and profiles~--- the analogs of~$D$  
and $F_s(\omega_s)$ in Eqs.~(\ref{eq:an}), (\ref{eq:ab})~--- 
and identification of a ``nonthermal attractor'' which is
often unique~\cite{Semikoz:1994zp, *Semikoz:1995rd,  Nowak:2012gd,  
  Berges:2014bba, PineiroOrioli:2015cpb, Chantesana:2018qsb,
  Schmied:2018mte, Semisalov:2021}.

In this paper extending  Ref.~\cite{Dmitriev:2023ipv} we
systematically study self-similar kinetics driven by
long-range\footnote{Note that electromagnetic interactions are
effectively short-range in plasma~\cite{pitaevskii2012physical,
  2013PhyU...56...49Z} due to Debye screening; cf.\ kinetics
of gravitational waves in Ref.~\cite{Galtier:2021ovg} and kinetics
  of growing cosmological structures in Refs.~\cite{Amin:2025sla,
    Amin:2025nxm}.} gravitational 
forces. To be precise, we consider highly populated, gravitationally
interacting bosonic gas in the vicinity of Bose-Einstein condensate
(Bose star). We simplify kinetic equation for this gas: ignore its
mean spatial inhomogeneities and describe two-particle  gravitational
scattering via approximate Landau
integral~\cite{Levkov:2018kau}, see
  also Refs.~\cite{pitaevskii2012physical, Chavanis:2020upb}. We account 
for particle  and energy exchanges with the condensate using
particle-absorbing (condensing) boundary conditions at low energies and an external
source heating the~gas.

We show that with these simplifications kinetic equation has a family
of self-similar solutions~(\ref{eq:an}), (\ref{eq:ab})~---  attractors
of kinetic evolution. Namely, if the  external source is absent, the
evolution inevitably approaches scaling regime with~${D=5/2}$, and if
the source is scale-symmetric,    self-similarity occurs at
different~$D$. We classify all self-similar solutions with finite
masses, energies, and  fluxes.

Our self-similar profiles~$F_s(\omega_s)$ include\footnote{We
  have found profiles with other,~${F_s \propto 
    \omega_s^{-1/3}}$, low-energy behavior. But they form smaller family with one
  parameter fixed and decay in time-dependent numerical  
  simulations. They are likely to be fine-tuned and unstable.}  thermal 
(Ray\-leigh–Je\-ans) low-energy tails~${F_s \propto \omega_s^{-1/2}}$  and 
cutoffs at high~$\omega_s$, see the sketch in
Fig.~\ref{fig:schematic}(b). Besides, they support finite particle
fluxes at the lowest~$\omega_s$ (``IR cutoff'' in the figure)
below which the particles are eaten by the condensate. This structure
is unusual: in the standard case of short-range interactions
the respective tails are nonthermal\footnote{They are often
close~\cite{Semikoz:1994zp, *Semikoz:1995rd, Semisalov:2021} to
Kolmogorov-Zakharov power-law cascades~\cite{Zakharov:1966,
  Nazarenko:2011, zakharov2012kolmogorov} describing 
particle or energy transport in the phase space. In our case, the
  only cascade~${F \propto \omega^{-1/3}}$ is likely to be unstable,
  see details in Sec.~\ref{sec:absence-power-law}.}
power-laws~\cite{Semikoz:1994zp,  *Semikoz:1995rd,
  Semisalov:2021}. Still, our self-similar solutions
are far from the true equilibrium, since all their parts
including the thermal tails evolve according to Eqs.~(\ref{eq:an}),~(\ref{eq:ab}).  

Most importantly, we observe that kinetic evolution remains
approximately self-similar even if scale symmetry of kinetic
equation gets explicitly broken by external conditions, e.g., by 
the arbitrarily chosen energy source. In this new\footnote{The closest
phenomenon is prescaling~\cite{Schmied:2018upn, Heller:2023mah}: nearly
self-similar evolution prior to reaching true~--- and unique~---
nonthermal attractor. In our case, self-similar solutions exist at
different~$D$ and adiabatic self-similarity means flow between
them.} regime of {\it adiabatic self-similarity}~\cite{Dmitriev:2023ipv} the evolution
follows the scaling law~(\ref{eq:an}), (\ref{eq:ab}) over long
epochs as its parameter~${D = D(t)}$ slowly 
drifts with time. We expect that all scale-breaking effects, even the
ones we ignored,  act likewise:
induce slow evolution in the solution space instead of ruining
self-similarity. This assumption is supported by observation of
self-similar behavior in microscopic (Sch\"odinger-Poisson)
simulations~\cite{Dmitriev:2023ipv}. 

We discover that adiabatic self-similarity is an efficient
analytical tool for non-equilibrium kinetics. One uses it by
computing a sequence of self-similar solutions with different~$D$
and then extracting slow time dependence of  the latter
parameter, say, from the conservation laws. An expansion
  constant in this approach is a relative contribution of
  scale-breaking effects estimated as
\begin{equation}
  \label{eq:41}
  d\ln D/d\ln(t - t_i)\ll 1\,.
\end{equation}
We consider two applications of the adiabatic method: condensation in
the presence of scale-breaking source and
growth~\cite{Dmitriev:2023ipv} of Bose star with fixed energy-mass
relation. In both cases we successfully describe nontrivial kinetic   
evolution up to few heuristic constants.

We believe that  adiabatic self-similarity is capable of
resolving~\cite{Dmitriev:2023ipv} a confusion  with long-time
  growth of Bose stars in the literature,
    cf.\ Ref.~\cite{Chavanis:2025qcg}. Simulations of structure 
formation~\cite{Schive:2014dra, *Schive:2014hza, *Liao:2024zkj,
  Mocz:2017wlg, Veltmaat:2018dfz, *Schwabe:2021jne, Schwabe:2020eac, 
  Eggemeier:2019jsu, Mina:2020eik, Amin:2025sla} indicate that
their masses~$M_{bs}$ increase on kinetic
timescales at first, but  then slow
down to almost a full halt once certain
``core-halo'' point is reached~\cite{Schive:2014dra, *Schive:2014hza,
  *Liao:2024zkj}. Different
studies report  contradicting 
``core-halo'' masses, cf.\ \cite{Schive:2014dra,   *Schive:2014hza,
  *Liao:2024zkj, Veltmaat:2018dfz, *Schwabe:2021jne}
and~\cite{Mocz:2017wlg, Nori:2020jzx, Mina:2020eik}, their
scatter~\cite{Chan:2021bja} and drift~\cite{Blum:2025aaa}. To add   
to this mess, long simulations  in artificial environments (gas-filled
boxes and hand-made miniclusters) confirm slow Bose star
  growth, but suggest different  laws:~${M_{bs} \propto 
  t^{1/2}}$ in Ref.~\cite{Levkov:2018kau},~$t^{1/4}$ in
Ref.~\cite{Chan:2022bkz} and~$t^{1/8}$ in
Refs.~\cite{Eggemeier:2019jsu, Chen:2020cef}.

Undoubtedly, the above slow-down of kinetic evolution 
signals of a ``nonthermal attractor'' reached by the gas-Bose-star
system; in Ref.~\cite{Dmitriev:2023ipv} we numerically confirmed its
self-similar structure.  The other effects~---
scatter of the data and variability of growth laws~---  can be  explained
by long-time drift of the scaling exponent~${D = D(t)}$ along with spread
of timescales in different simulations. 

This paper is organized as follows. In Sec.~\ref{sec:kinetic} we
introduce simplified kinetic equation for gravitational scattering and
discuss its scale symmetry.  Numerical example of kinetic
evolution approaching self-similar attractor is given in 
Sec.~\ref{sec:attr-self-simil}. Section~\ref{sec:absence-power-law}
  reviews stationary power-law cascades in gravitational kinetics. 
  Scaling solutions are studied in Sec.~\ref{sec:self-simil-solut},
  and adiabatic self-similarity is introduced in
  Sec.~\ref{sec:compare}. Section~\ref{sec:discussion} contains
  concluding remarks.


\section{Scale symmetry of gravitational kinetics} 
\label{sec:kinetic}

Consider an overpopulated gas of nonrelativistic bosons with
two-particle gravitational interactions~\cite{Levkov:2018kau}. Three
simplifications make its kinetic equation 
scale-invariant. First, ignore spatial inhomogeneity of the gas and
its collective gravitational field,
\begin{equation}
  \label{eq:5}
  f_{\bm{p}}(t, \, \bm{x}) = f_{\boldsymbol{p}}(t)
  \qquad \mbox{and}\qquad \bar{U}(t,\, \bm{x}) = 0\,,
\end{equation}
where~${f}_{\bm{p}}$ is the phase-space density and~$\bm{p}$
is particle momentum. Second, assume large occupation numbers   
\begin{equation}
  \label{eq:6}
  f_{\bm{p}}\gg 1\,,
\end{equation}
i.e.\ a non-thermal state below the condensation point. Third, use
Landau (long-range) approximation for gravitational
scattering~\cite{pitaevskii2012physical}.

These approximations give kinetic equation\footnote{Of the same
form as equation for random waves in the Coulomb-interacting
plasma~\cite{pitaevskii2012physical, zakharov2012kolmogorov}, but
with electric charge replacement~${e \to
  m\sqrt{G}}$~\cite{Levkov:2018kau, Chavanis:2020upb}.}~\cite{Levkov:2018kau}
\begin{equation}
  \label{eq:LE}
 \partial_{t}f_{\bm{p}} = \mathrm{St}\, f_{\bm{p}} \equiv - \partial_{p_j}s_j(\bm{p}) \,,
\end{equation}
with scattering integral~${\mathrm{St}\, f_{\bm{p}}}$ and Landau flux 
\begin{equation}
  \label{eq:si}
  s_i(\bm{p}) = \frac{G^2m^4 \Lambda}{4\pi^2}
  \int \frac{d^3 \bm{q}}{|\bm{u}|}\;  {\cal P}_{ij}\left[
    f_{\bm{p}}^2\partial_{q_j}f_{\bm{q}} - f_{\bm{q}}^2\partial_{p_j}f_{\bm{p}} \right]
\end{equation}
describing gravitational collisions of particles with
momenta $\bm{p}$ and $\bm{q}$; we introduced their relative 
velocity~${\bm{u} = (\bm{p} - \bm{q})/m}$ and a projector ${{\cal 
    P}_{ij} = \delta_{ij} - u_i   u_j/|\bm{u}|^2}$ orthogonal
to~$\bm{u}$. The strength of  gravitational scat\-te\-ring in
Eq.~(\ref{eq:si}) is controlled by the particle mass~$m$, gravitational
constant~$G$, and a Coulomb logarithm ${\Lambda =   \ln(p_0 R) \gtrsim
  1}$ of reference particle momentum~$p_0$ and
spatial gas extent~$R$.

Now, the simplified kinetic equation~(\ref{eq:LE}), (\ref{eq:si}) has scale symmetry 
\begin{equation}
  \label{eq:7}
{f \to a f}\,,  \quad {\bm{p} \to \bm{p}/b}, \quad  t-t_i \to (t-t_i)/a^2
\end{equation}
with constants~$a$,~$b$ and time origin $t_i$. In other words, the
function
\[
  f'_{\bm{p}}(t-t_i) = a f_{b\bm{p}}(a^2(t-t_i))
\]
satisfies the equation if $f_{\bm{p}}(t)$ does.

We will argue in Sec.~\ref{sec:self-simil-solut} that
the symmetry~(\ref{eq:7}) guarantees existence of 
self-similar kinetic solutions~(\ref{eq:an}),~(\ref{eq:ab}). In
reality, it  is broken by free-streaming effects and mean 
gravitational field  of the inhomogeneous gas, as well as by the
short-range and low-occupancy~($1/f$) corrections to the scattering
integral. However, soon  we will see that self-similar
solutions are dynamical attractors. Physically relevant
scale-breaking effects do not destroy them but induce 
slow drift of their parameters.

Before proceeding, let us further simplify Eqs.~(\ref{eq:LE}),
(\ref{eq:si}) by suggesting that the gas is isotropic,
i.e.\ $f_{\bm{p}}$ depends only on~$\bm{p}^2$. In this case it is
convenient to work with the distribution of particles over
energies~$\omega$,
\begin{equation}
  \label{eq:9}
  F(t,\, \omega) \equiv \frac{dN}{d\omega} = \frac{mV_R}{2\pi^2}\, p
  f_{\bm{p}} \,, \qquad \omega = \frac{\bm{p}^2}{2m}\,,
\end{equation}
where $V_R$ is the spatial volume of the gas. Kinetic
equation~(\ref{eq:LE}), (\ref{eq:si}) takes the form (see 
App.~\ref{sec:KE_hsse} and Ref.\footnote{As compared
to~\cite{Levkov:2018kau}, we changed normalizations of basic integrals:
${A \to mV_R^2 A/(4\pi^4)}$ and~${B \to V_R
    B/(2\pi^2)}$.}~\cite{Levkov:2018kau}):  
\begin{subequations}
  \label{eq:14}
  \begin{equation}
    \label{eq:2}
    \partial_t F = \mathrm{St}\, F\,, \qquad
    \mathrm{St}\, F =  - \partial_\omega {\cal J}_N = - \omega^{-1}
    \partial_\omega {\cal J}_E\,,
  \end{equation}
where we introduced particle and energy fluxes in the~$\omega$
space,
\begin{equation}
 \label{eq:8}
  {\cal J}_{N} = \partial_{\omega} W(\omega) \quad\mbox{and}\quad 
  {\cal J}_{E} = \omega^2  \partial_{\omega} \left( W(\omega)/\omega\right)\,,
\end{equation}
and preflux
\begin{equation}
  \label{eq:10}
  W(\omega) = W_0\left( B C
  - A F \right)\,, \qquad W_0 \equiv \frac{(2\omega_0)^3}{t_{\mathrm{rel}} N_0^2} 
\end{equation}
involving three basic integrals
\begin{align}
  \label{eq:ABC}
  & A(\omega) = 
  \int\limits_0^{\infty} d\omega' \; \frac{\mathrm{min}^{3/2}(\omega,
    \omega')}{3\omega' \omega^{1/2}}\; F^2(\omega')\,,\\ \notag
  & B(\omega) = \int\limits_0^{\omega} d\omega' F(\omega')\,, \;\;
  \mbox{and} \;\;
  C(\omega) = \int\limits_{\omega}^{\infty} \frac{d\omega'}{2\omega'} \;
  F^2(\omega')\,.
\end{align}
\end{subequations}
Note that the last equality in Eq.~(\ref{eq:2}) is a trivial
consequence of the definitions~(\ref{eq:8}). In Eq.~(\ref{eq:10}) we
used reference parameters:  typical particle energy~${\omega_0 =
  p_0^2/(2m)}$, multiplicity~$N_0$, and
theoretical relaxation time 
\begin{equation}
  \label{eq:36}
  t_{\mathrm{rel}} = \frac{\omega_0^3 V_R^2}{\pi^3 N_0^2 m^2 G^2 \Lambda} \sim t_{gr}\,.
\end{equation}
The latter roughly coincides\footnote{Strictly speaking,
$t_{\mathrm{rel}} \equiv 3 t_{gr}/(2b\sqrt{2})$, where~$b\sim
0.6\div 0.9$~\cite{Levkov:2018kau}.}~\cite{Levkov:2018kau}  the with the time~$t_{gr}$ of
kinetic Bose star formation in simulations.  

Note that gas isotropy neither favors nor hinders self-similar
  dynamics: we consider this case as simple and physically
motivated. In terms of~$F$, scaling transformations~(\ref{eq:7})
look like
\begin{equation}
  \label{eq:11}
  F \to F'(t-t_i,\, \omega) = \frac{a}{b}\, F(a^2(t-t_i),\, b^2 \omega)\,.
\end{equation}
They leave the kinetic equation~(\ref{eq:2}) invariant.

\begin{figure}[t]
  \centerline{\includegraphics{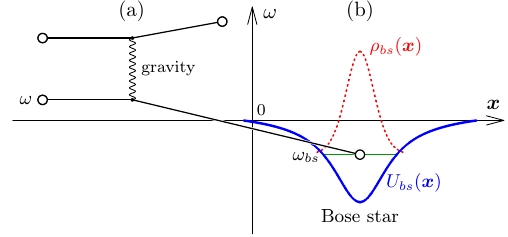}}
  \caption{(a)~A diagram for particle condensation onto the Bose
    star. (b)~Bose star gravitational field~$U_{bs}(\bm{x})$ and its
    mass density~$\rho_{bs}(\bm{x})$.}
  \label{fig:condensation_dia}
\end{figure}

Next, we add Bose-Einstein condensate to the system. In gravity,
condensed particles create potential well~$U_{bs}(\bm{x}) < 0$ and
simultaneously occupy its lowest energy level~${\omega_{bs} <
0}$. As a result, they form gravitationally self-bound object with
localized mass density~$\rho_{bs}(\bm{x})$ called Bose
star~\cite{Tkachev:1986tr}, see Fig.~\ref{fig:condensation_dia}(b).

We will not study the Bose star itself, only its effect on the
surrounding gas (see,
however,~\cite{Chavanis:2011zi, Eby:2015hsq, Levkov:2016rkk,
  Eby:2019ntd, Dmitriev:2021utv, Visinelli:2021uve, Chan:2023crj,
  Salasnich:2025nrt}). On the one 
hand, condensation of particles onto the star means their
transitions to the level~${\omega_{bs}<0}$ and hence nonzero particle
flux through the  boundary~${\omega=0}$ of the gas energy
space. An example of  such process is given in
Fig.~\ref{fig:condensation_dia}(a): a particle loses its energy
in gravitational collision and appends to the 
condensate. We  imitate condensation\footnote{An opposite
process knocking the particles off the level~$\omega_{bs}$
evaporates the condensate~\cite{Chan:2022bkz}. Our effective
term in the equation describes difference between condensation and 
evaporation.} with  effective absorbtion term in the kinetic
equation. Since gravitational scattering is more efficient at 
lower energy transfers~\cite{pitaevskii2012physical}, the Bose star
mainly eats low-$\omega$ particles~\cite{Chan:2022bkz}. Accordingly,
our  absorbtion term will act at small~$\omega
\lesssim\omega_{\mathrm{IR}}$~\cite{Dmitriev:2023ipv}. 

On the other hand, conservation of total energy in the gas-Bose-star
system implies that condensation heats the gas, i.e.\ raises its
energy. For example, condensing particle in
Fig.~\ref{fig:condensation_dia}(a) transfers its energy
excess~${\omega - \omega_s > 0}$ to the collision partner remaining
in the gas. Since heating is essential~\cite{Eggemeier:2019jsu,
  Dmitriev:2023ipv}, we  mimic  it with energy
source~$J_{\mathrm{ext}}$ in the kinetic equation. 

These two physical motivations suggest kinetic equation
\begin{equation} 
  \label{eq:13}
    \partial_t F = \mathrm{St}\, F - \mu(\omega) F + J_{\mathrm{ext}}(t,\, \omega)\,,
\end{equation}
modeling particle and energy exchanges with the condensate. In
particular, the 
sponge~$\mu(\omega)$ localized at~$\omega \lesssim
\omega_{\mathrm{IR}}$ absorbs particles and the
source~$J_{\mathrm{ext}}$ pumps energy into the gas.

Skeptics may say  that Eq.~(\ref{eq:13}) is valid at the
qualitative level. More detailed description would include Bose star
gravitational field and  provide~$F$--dependent expressions for~$\mu$
and~$J_{\mathrm{ext}}$ on the basis of proper collision integral,
cf.\ Refs.~\cite{Chan:2022bkz, Dmitriev:2023jnu}. However, we use
Eq.~(\ref{eq:13}) as a  
convenient testing tool for self-similar kinetics. Besides,
we  do not expect neglected terms to move evolution too far away
from  self-similar attractors: in Sec.~\ref{sec:compare} we will
see that reasonable heuristic choices of~$\mu$ and~$J_{\mathrm{ext}}$
give very good results, see also Ref.~\cite{Dmitriev:2023ipv}.

In general, the sponge~$\mu(\omega)$ and the
source~$J_{\mathrm{ext}}(t,\, \omega)$ break the scale
symmetry~(\ref{eq:11}). We exploit this feature in two
ways. First, when studying exact self-similar dynamics we
restore  the subgroup of symmetry with~${b = a^{1-2/D}}$ at
  constant~$D$. To this end we consider sources satisfying
\begin{equation}
  \label{eq:3}
  J_{\mathrm{ext}}(t-t_i,\, \omega) = a^{2+2/D} \, J_{\mathrm{ext}}
  (a^2 (t-t_i),\, a^{2-4/D}\omega)
\end{equation}
and place an infinitely strong sponge at $\omega \approx 0$. The
latter imitates Bose star that swallows all zero-energy particles
  and ignores the rest of them. It is equivalent to the boundary condition
\begin{equation}
  \label{eq:1}
  {\cal J}_E = 0 \quad \mbox{and} \quad {\cal J}_N \ne 0  \quad
  \mbox{at} \quad \omega=0\,.
\end{equation}
 One can check that
equation~(\ref{eq:13}) with scattering
  integral~(\ref{eq:14}),~${\mu=0}$, boundary 
condition~(\ref{eq:1}), and the source~\eqref{eq:3} is invariant under 
the transformation (\ref{eq:11}) with~${b = a^{1-2/D}}$. This is
  enough for the existence of self-similar solutions.

Second, in Sec.~\ref{sec:compare} we will deliberately consider
generic functions~$\mu = \mu(\omega)$ and~$J_{\mathrm{ext}} =
J_{\mathrm{ext}}(t,\, \omega)$ as an origin of scale-breaking
  effects leading to adiabatic self-similarity. 


\section{Attracting to self-similar profiles}
\label{sec:attr-self-simil}

Let us numerically evolve time-dependent kinetic
equation~(\ref{eq:13}) at~${J_{\mathrm{ext}} = 0}$. To this end we
place the sponge at tiny~${\omega\lesssim \omega_{\mathrm{IR}}
  = 2\cdot 10^{-3}\, \omega_0}$,
\begin{equation}
  \label{eq:17}
  \mu = \mu_0\,\vartheta[(\omega_{\mathrm{IR}}-\omega)/\sigma_{\mathrm{IR}}]\,,
\end{equation}
where~${\vartheta(x) = [1 + \mathrm{th}\, ( 2x+x^3 )]/2}$ is a smoothed
theta-function,~${\mu_0 = 2\cdot 10^6/t_{\mathrm{rel}}}$, and~${\sigma_{\mathrm{IR}} =
\omega_{\mathrm{IR}}/40}$. We also adopt dimensionless units: measure $t$,
$F$, and $\omega$ in terms of the relaxation time~$t_{\mathrm{rel}}$,
reference particle number~$N_0$, and energy~$2\omega_0$,
\begin{equation}
  \label{eq:12}
  \tau \equiv t/t_{\mathrm{rel}}\,,\qquad \tilde{F} = 2\omega_0
  F/N_0\,, \qquad \tilde{\omega} = \omega/(2\omega_0)\,,
\end{equation}
thus removing these constants from Eqs.~(\ref{eq:14}), (\ref{eq:13}). 

\begin{figure}
  \centerline{\includegraphics{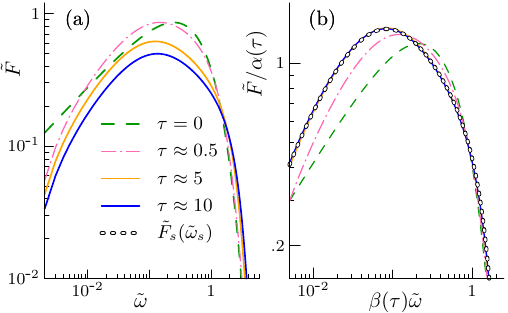}}
  \caption{(a)~Numerical solution of the kinetic equation (\ref{eq:13})
    with the sponge~(\ref{eq:17}),~$J_{\mathrm{ext}}=0$,  
    and Gaussian initial data~\eqref{eq:16}. Lines 
    show~$\tilde{F}$ as a function of~$\tilde{\omega}$ at 
    different~$\tau\equiv t/t_{gr}$. (b)~The same solution rescaled
    by~$\alpha(\tau)$ and~$\beta(\tau)$ from Eq.~(\ref{eq:15})
    with~${D = 5/2}$ and~$\tau_i \approx -2.3$. Chain points display
    self-similar profile~$F_s(\omega_s)$ satisfying Eq.~\eqref{eq:4}
    with $J_{\mathrm{ext},s}=0$.}
  \label{fig_attract_kin}
\end{figure}

We start at $\tau=0$ from the Gaussian initial profile
\begin{equation}
  \label{eq:16}
  f_{\bm{p}} \propto \mathrm{e}^{-(\bm{p}/p_0)^2}\quad
  \mbox{and} \quad  \tilde{F}(0, \tilde{\omega}) = c_N\,
  \tilde{\omega}^{1/2}\, \mathrm{e}^{-2\tilde{\omega}}
\end{equation}
with multiplicity~${N(0) =  2^{-5/2} c_N  N_0 \sqrt{\pi}}$ and
  typical energy ${\omega_0 \equiv p_0^2/(2m)}$ hidden in the
  tildes. Taking ${c_N = 2^{3/2}}$, we numerically evolve kinetic
equation~(\ref{eq:13})  
with the sponge~(\ref{eq:17}), scattering integral~\eqref{eq:14},
and ${J_{\mathrm{ext}}=0}$; see App.~\ref{sec:numer-solut-time} for
numerical details. The solution $\tilde{F}(\tau,\, \tilde{\omega})$ is
visualized in Fig.~\ref{fig_attract_kin}(a). After a period of
  rapid adjustment at~$0 \leq\tau\lesssim 0.5$ it basically retains
  its shape with a noticeably decreasing amplitude due to particle
  absorption.  This evolution is remarkably slow: thermalization
is still far away  at~${\tau \equiv t/t_{\mathrm{rel}} \sim 10}$.

Now, we observe that the late-time kinetic evolution in
Fig.~\ref{fig_attract_kin}(a) reduces to time-dependent
rescaling~\eqref{eq:an} of a single self-similar
profile~$F_s(\omega_s)$. Indeed, let us divide~$F$ and
multiply~$\omega$ by time-dependent coefficients~(\ref{eq:ab}),
\begin{equation}
  \label{eq:15}
  \alpha(\tau) =  (\tau - \tau_i)^{-1/D}\quad \mbox{and} \quad \beta(\tau)  = (\tau -
  \tau_i)^{2/D-1}\,.
\end{equation}
In Fig.~\ref{fig_attract_kin}(b) this makes
all the graphs  with~${\tau\geq 5}$ (solid lines) merge into a single
curve because the scaling parameters~${D = 5/2}$ and~${\tau_i \approx
  -2.3}$ were selected correctly; the latter choice will be explained below.

Numerical experiment in Fig.~\ref{fig_attract_kin} strongly suggests
that self-similar solution~(\ref{eq:an}), \eqref{eq:ab}
with~$D = 5/2$ is a kinetic attractor. Indeed, the distribution
  function in Fig.~\ref{fig_attract_kin}(b) essentially changes its
  form during the first relaxation interval~${\tau \lesssim 1}$
  from the Gaussian initial hat to scaling profile
  $F_s(\omega_s)$, cf.\ the dashed and solid lines.  We
solved the kinetic equation with essentially different initial
  condition~${\tilde{F}(0,\,\tilde{\omega}) =
  \delta(\tilde{\omega}-1)}$ and obtained similar result: at~${\tau
  \gtrsim 8}$ the distribution function approached the same scaling
solution as in Fig.~\ref{fig_attract_kin}(b), albeit with
parameters~${D = 5/2}$ and~${\tau_i \approx 0.15}$.

One can easily explain, why scaling dimension~$D=5/2$ invariably emerges in 
evolutions with~$J_{\mathrm{ext}}=0$. Indeed, zero source
  makes the gas energy almost time--independent, since
the only nonconservative term~--- the sponge~\eqref{eq:17}~---  eats
only particles with $\omega\approx 0$. On the other hand, the energy of  
self-similar solution~(\ref{eq:an}), (\ref{eq:ab}) equals
\begin{equation}
  \label{eq:21}
  E = \int\limits_0^\infty \omega d\omega \,  F(t,\, \omega)
  \propto (t-t_i)^{2-5/D}\,.
\end{equation}
It is conserved precisely at~${D = 5/2}$, hence the choice of this
  parameter\footnote{Solutions with generic~$D$ will be 
    classified in Sec. \ref{sec:self_eq}.} in Fig.~\ref{fig_attract_kin}(b).

Note that the value of~$\tau_i$~--- the time shift of 
  the scaling attractor~--- strongly depends on the initial
  distribution function and cannot be fixed by the conservation laws. We
  determine it by fitting the numerical data.

\begin{figure}
  \centerline{\includegraphics{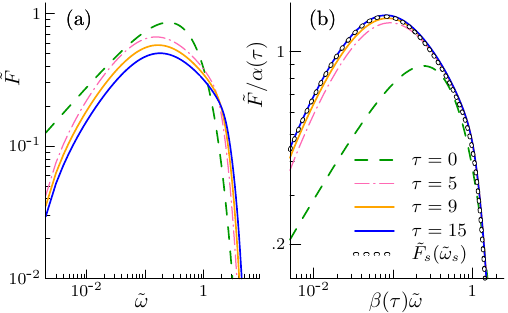}}
  \caption{(a)~Kinetic evolution with the sponge~(\ref{eq:17})
      and a nonzero energy source~(\ref{eq:19}) starting from the
      Gaussian initial profile~\eqref{eq:16}. Graphs show the
      distribution function $\tilde{F}(\tau,\, \tilde{\omega})$
      at different~$\tau$. (b)~The same graphs rescaled with~$\alpha(\tau)$ 
      and~$\beta(\tau)$ in Eq.~(\ref{eq:15}), where~${D = 2.8}$
      and~${\tau_i = -1.1}$. Chain points give self-similar  
      profile~$\tilde{F}_s(\tilde{\omega}_s)$ extracted from
      Eq.~\eqref{eq:4}.} 
  \label{fig_attract_kinJ}
\end{figure}

By contrast, we expect appearance of scaling dynamics with~${D\ne
  5/2}$ if~$J_{\mathrm{ext}}$ is
  nonzero. Figure~\ref{fig_attract_kinJ} shows kinetic evolution  of
  the same initial data~(\ref{eq:16}) with
  \begin{equation}
  \label{eq:19}
  J_{\mathrm{ext}}(\tau,\, \omega) = \frac{J_0}{t_{\mathrm{rel}}}\;
  \frac{\alpha^3(\tau) \beta(\tau)}{\mathrm{ch}^{2} 
    \left[\omega \beta(\tau)/\sigma - \omega_1 \right]}\,,
\end{equation}
  where~$\mathrm{ch}^2(x)$ is a hyperbolic cosine squared, while~$\alpha$
  and~$\beta$ are given by Eqs.~(\ref{eq:15}) with ${D = 
  2.8}$ and~$\tau_i = -1$; we use ${J_0 = 3N_0/(20 \omega_0)}$, ${\sigma =
0.4\,\omega_0}$, and ${\omega_1 = 5}$.  Rescaling this evolution in 
Fig.~\ref{fig_attract_kinJ}(b), we find out
that it becomes self-similar  with~${D = 2.8}$  and~${\tau_i =
  -1.1}$ at~${\tau \gtrsim 9}$. Thus, the solution~(\ref{eq:an}),
(\ref{eq:ab}) is still a dynamical attractor at~${J_{\mathrm{ext}}\ne 0}$,
but it has~${D\ne 5/2}$.

We will show shortly that exact self-similar dynamics arises
  in Fig.~\ref{fig_attract_kinJ} at large~$\tau$ only because the
  source~(\ref{eq:19}) preserves restricted scale
  symmetry~(\ref{eq:11}), i.e., satisfies Eq.~\eqref{eq:3}
  with~${D = 2.8}$. For generic sources, self-similar
evolution is attractive but approximate.

\section{Useless power-law cascades}
\label{sec:absence-power-law}

\begin{figure}
  \centerline{\includegraphics{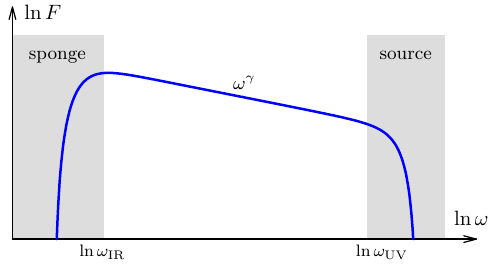}}
  \caption{Typical Kolmogorov-Zakharov cascade (not to scale).}
  \label{fig:cascade}
\end{figure}

In ordinary short-range kinetics, the setup~(\ref{eq:13}) with an
absorbing sponge and a source sets the stage for cascading
Kolmogorov-Zakharov  solutions~\cite{Zakharov:1966, Nazarenko:2011, 
  zakharov2012kolmogorov}. The latter have time-independent
  power-law profiles
\begin{equation}
  \label{eq:18}
  F(\omega) = F_0 \, \omega^{\gamma} 
\end{equation}
and describe transport of conserved quantities
between infrared and ultraviolet cutoffs~$\omega_{\mathrm{IR}}$
and~$\omega_{\mathrm{UV}}$~--- the sponge/source positions,
see Fig.~\ref{fig:cascade}. The cascade~(\ref{eq:18}) is 
local~\cite{zakharov2012kolmogorov}, i.e.\ exists as an independent
solution, if its scattering integral at $\omega_{\mathrm{IR}} \ll
\omega \ll \omega_{\mathrm{UV}} $ receives negligible contributions 
from the regions near the cutoffs~$\omega \approx \omega_{\mathrm{IR}}$
and~$\omega_{\mathrm{UV}}$.
 
Let us demonstrate that in long-range kinetics cascading solution 
exists but has strange properties hampering its physical
interpretation. Substituting Eq.~(\ref{eq:18}) with the cutoffs 
into collision integral~(\ref{eq:14}), we find, 
\begin{multline}
  \label{eq:StFcascade}
  \mathrm{St}\, F = W_0 F_0^3 \omega^{3 \gamma-1} \Bigg[ \frac{3 (1
      + 2 \gamma) (1 + 3 \gamma)}{4 (1 + \gamma) (3 + 4 \gamma)}  + 
    O\left(\frac{\omega_{\mathrm{IR}}}{\omega}\right)^{\gamma+1}\\
    + O(\omega_{\mathrm{IR}}/\omega)^{2\gamma + 3/2}
    + O(\omega/\omega_{\mathrm{UV}})^{-2\gamma} \Bigg]\,,
\end{multline}
where the last three terms estimate contributions
from $\omega_{\mathrm{IR}}$ and~$\omega_{\mathrm{UV}}$. The latter are
irrelevant and the cascade is local if\footnote{The
prefactors of the boundary terms vanish at~${\gamma=1/2}$ which
corresponds to regular phase-space density~$f_{\bm{p}} \propto F/p$
at~${\bm{p}=0}$, i.e.\ no condensate.}~${-3/4 < \gamma < 0}$.  

Zeros of the scattering integral~(\ref{eq:StFcascade})
\begin{equation}
  \gamma = -1/2 \qquad \mbox{and} \qquad \gamma = -1/3
\end{equation}
correspond to stationary solutions of the kinetic equation. 
Here the first option is Rayleigh–Jeans distribution,
\begin{equation}
  \mbox{thermal:} \quad f_{\boldsymbol{p}} \propto \bm{p}^{-2}\,, \quad
  \mbox{and hence} \quad F \propto \omega^{-1/2}\,,
\end{equation}
i.e.\ thermal equilibrium with zero fluxes.

The second option $F \propto \omega^{-1/3}$ is more involved. Its
fluxes~--- primitives~(\ref{eq:2}) of the scattering integral~---
are independent of~$\omega$:
\begin{equation}
  \label{eq:20}
  {\cal J}_N = 0 \;\; \mbox{and} \;\; {\cal J}_E = -\frac{9}{40}
  W_0 F_0^3 < 0 \;\; \mbox{at} \;\; \gamma = -1/3\,.
\end{equation}
Regrettably, this cascade describes neither gas condensation nor
heating of its thermal tail which require~${{\cal J}_N<0}$ and~${{\cal
    J}_E > 0}$, respectively. As a consequence, the
solution~(\ref{eq:20}) does not appear in time-dependent kinetic
simulations which usually move in the opposite direction. The
only situations where it may be of any use are
counter-thermalizing processes like condensate evaporation in
Ref.~\cite{Chan:2022bkz}.

Despite problems with physical interpretation, in
Sec.~\ref{sec:self_eq} we upgrade the~$\omega^{-1/3}$ cascade to  
self-similar solutions which~--- alas~--- happen to be unstable.

\section{Self-similar solutions}
\label{sec:self-simil-solut}

\subsection{The Ansatz}
\label{sec:ansatz}

One notices that~(\ref{eq:an}), (\ref{eq:ab}) is the most
general Ansatz invariant under the subgroup\footnote{The only
function~${F \propto \omega^{1/2}/(t-t_i)^{1/2}}$ invariant under
the full group is a very special solution with infinite
  charges and uniform phase-space density~${f_{\bm{p}}
    \propto (t - t_i)^{-1/2}}$.} of scaling
transformations~(\ref{eq:11}) with~${b = a^{1-2/D}}$, where~$D$ is a
constant. Indeed,
\begin{equation}
  \label{eq:22}
  \omega_s = \beta(\tau) \omega
\end{equation}
is an invariant by itself and~$\alpha(\tau)$ tranforms like~$F$;
as before,~${\tau \equiv t/t_{\mathrm{rel}}}$. By
Coleman's theorem~\cite{Rubakov:2002fi}, this guarantees that the
Ansatz~(\ref{eq:an}), (\ref{eq:ab}) passes the kinetic 
equation~(\ref{eq:13}),~\eqref{eq:14} with~${\mu=0}$ and
scale-invariant source~\eqref{eq:3}. 

Substituting~${F(t,\,\omega) = \alpha(\tau) F_{s}(\omega_s)}$ and
Eqs.~\eqref{eq:22}, \eqref{eq:15}, we indeed arrive at time-independent
equation for~$F_s(\omega_s)$,
\begin{equation}
  \label{eq:4}
(2/D-1) \, \omega_s \partial_{\omega_s}F_s - F_s/D =
  t_{\mathrm{rel}}\,\mathrm{St}\,  F_s + J_{\mathrm{ext},s}(\omega_s)\,,
\end{equation}
where the sponge was traded for scale-invariant boundary
condition~\eqref{eq:1} and we introduced the rescaled source
\begin{equation}
  \label{eq:23}
  J_{\mathrm{ext},s}(\omega_s) \equiv t_{\mathrm{rel}}\, J_{\mathrm{ext}}(t,\,
  \omega)/[\alpha^3(\tau) \beta(\tau)]
\end{equation}
which depends only on~$\omega_s$ if~$J_{\mathrm{ext}}$ has the 
symmetry~\eqref{eq:3}. Note that the rescaled collision integral $\mathrm{St}\,
F_s$ in Eq.~\eqref{eq:4} and rescaled fluxes~${\cal J}_{N,s}$
and~${\cal J}_{E, s}$ in the boundary condition~\eqref{eq:1} for~$F_s$
are given by the same expressions~\eqref{eq:14} as before, but
with~$F_s$ and~$\omega_s$ replacing~$F$ and~$\omega$. 

\begin{figure}
  \unitlength=1mm
  \centerline{\begin{picture}(84, 53)
      \put(0,0){\includegraphics{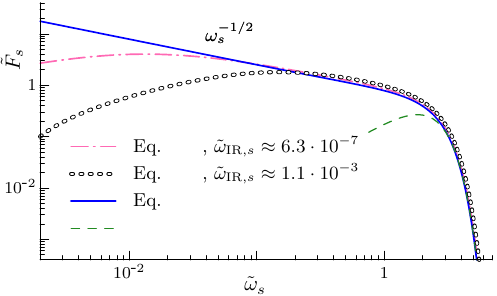}}
      \put(28,24.2){(\ref{eq:25})}
      \put(28,19.5){(\ref{eq:25})}
      \put(28,15.1){(\ref{eq:1})}
      \put(22.8,10.5){high-$\omega_s$ asymptotics (\ref{eq:30})}
  \end{picture}}
  \caption{Self-similar profiles satisfying Eq.~(\ref{eq:4})
      with~${J_{\mathrm{ext},s} = 0}$, ${D=5/2}$, and one of the boundary
      conditions: requirement~\eqref{eq:1} at~${\omega_s=0}$ (solid line) or 
      Eq.~\eqref{eq:25} at~${\omega_s = \omega_{\mathrm{IR},s}}$
      (chain points, dash-dotted line). We use dimensionless
      units~(\ref{eq:12}): ${\tilde{F}_s 
        \equiv 2\omega_0 F_s/N_0}$ and~${\tilde{\omega}_s \equiv
          \omega_s/(2\omega_0)}$. All solutions have~${{\cal J}_{N,s}(0) = -
          N_0/(2t_{\mathrm{rel}})}$. The chain point graph is repeated in
        Fig.~\ref{fig_attract_kin}(b) after rescaling~\eqref{eq:24}
        with~${b\approx 0.75}$.}
  \label{fig:Fs_D2.5}
\end{figure}
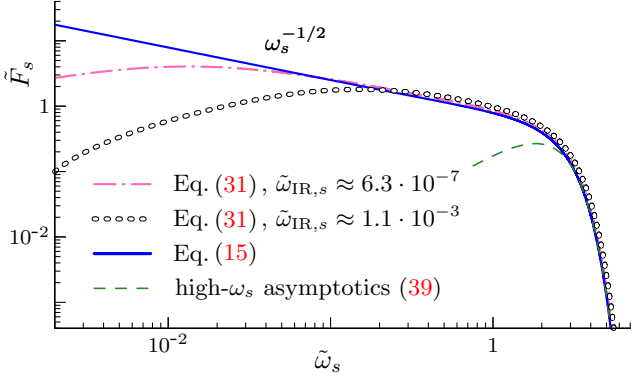

We obtained a well-posed boundary value problem \eqref{eq:1},
(\ref{eq:4}) for the profiles~$F_s(\omega_s)$ with finite 
multiplicities and energies. It gives one solution for
every ${J_{\mathrm{ext},s}(\omega_s) \ne 0}$ and~${D\ne 5/2}$. On the  
other hand, at~${J_{\mathrm{ext},s} = 0}$ and~${D=5/2}$ the same
problem has residual symmetry 
\begin{equation}
  \label{eq:24}
  F_s \to F_s'(\omega_s) = \frac{1}{b}\,  F_s(b^2 \omega_s)
\end{equation}
generating a family of 
solutions. We parametrize them by rescaled particle flux~${\cal
  J}_{N,s}$ at~${\omega_{s}=0}$ that, if negative, indicates
  condensation. Note that the original
  flux is   time-dependent: ${{\cal J}_N(t,\, 0) =
    \alpha^3(\tau) {\cal J}_{N,s}(0)}$, see Eqs.~\eqref{eq:an},~\eqref{eq:14}. 
  
Numerical profile~$F_s(\omega_s)$ satisfying Eq.~(\ref{eq:4})
with ${J_{\mathrm{ext},s}=0}$, ${D = 5/2}$, and boundary
condition~\eqref{eq:1} is plotted in Fig.~\ref{fig:Fs_D2.5}
by solid line; see App.~\ref{sec:self-simil-prof} for 
numerical method. This function has all characteristics previewed
in Fig.~\ref{fig:schematic}(b): thermal low-energy tail~${F_s \propto
\omega_s^{-1/2}}$ and exponential falloff as~${\omega_s \to +\infty}$.
 
Notably, at low~$\omega_s$ the above solution for~$F_s$
  essentially departs from the attractor profile in
time-dependent simulation of Fig.~\ref{fig_attract_kin}(b). This is
the effect of a finite-width sponge~(\ref{eq:17}) deforming the
simulation result even 
at~$\omega \gg \omega_{\mathrm{IR}}$. Indeed, let us mimic the sponge
by replacing Eq.~\eqref{eq:1} with the Dirichlet boundary condition
\begin{equation}
  \label{eq:25}
  F_s = 0 \qquad \mbox{at} \qquad \omega_s \leq \omega_{\mathrm{IR},s}\,.
\end{equation}
That changes the infrared tail of the profile, see the
chain points and dash-dotted line in Fig.~\ref{fig:Fs_D2.5}.
In particular, the chain-point graph rescaled by Eq.~\eqref{eq:24}
with~${b\approx 0.75}$ correctly reproduces\footnote{In 
  other numerical experiments
  we evolved time-dependent kinetic equation with
self-similar sponges, e.g.~${F=0}$ at ${\omega \leq
  \omega_{\mathrm{IR},s}/\beta(\tau)}$. Those evolutions
approached the profile solutions~$F_s(\omega_s)$ with the
same~$\omega_{\mathrm{IR},s}$.} the late-time result of 
simulation in Fig.~\ref{fig_attract_kin}(b).
 
The above exercise demonstrates that low-energy tails
of self-similar solutions can be strongly deformed at~${\omega_s \gg
  \omega_{\mathrm{IR},s}}$ by sponges of width~${\omega_{\mathrm{IR},s}}$. 
But in the limit $\omega_{\mathrm{IR},s} \to 0$ when the boundary 
condition~\eqref{eq:1} applies, universal thermal behavior
$F_s\propto \omega_{s}^{-1/2}$ emerges at low~$\omega_s$.


\subsection{Properties of scaling solutions}
\label{sec:self_eq}

\begin{figure}
  \centerline{\includegraphics[width=8.6cm]{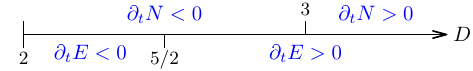}}
  \caption{Self-similar behaviors at different~$D$.} 
  \label{fig:ENranges}
\end{figure}

Self-similar solutions have time-dependent energies and multiplicities
and hence describe non-isolated gas. Indeed, the
Ansatz~(\ref{eq:an}),~\eqref{eq:15} gives  
\begin{equation}
  \label{eq:26}
  E = E_s (\tau - \tau_i)^{k_E} \,,\;\;
  N = \int\limits_{0}^{\infty} d\omega\,  F = N_s (\tau -
  \tau_i)^{k_N} 
\end{equation}
where Eq.~\eqref{eq:21} was used, we defined powers
\begin{equation}
 \label{eq:k}
k_E =  2-5/D\,, \quad  k_N = 1 - 3/D\,, \quad 3 k_E - 5k_N = 1\,,
\end{equation}
and total integrals
\begin{equation}
\label{eq:ENs}
E_s=\int_{0}^{\infty}\omega_sF_s(\omega_s)\, d\omega_s \,,\quad 
N_s=\int_{0}^{\infty}F_s(\omega_s)\,d\omega_s\,.
\end{equation}
Below we focus on solutions with finite~$E_s$ and~$N_s$.

We see that the scaling dimension~$D$ controls evolution of charges
and thus determines physical properties of solutions, see
Fig.~\ref{fig:ENranges}. In particular,~$E$ is conserved at~${D = 
  5/2}$ and~$N$ is at~${D = 3}$. Between these values at~${5/2 < D <
  3}$ self-similar evolutions  can describe particle condensation onto
the Bose star:~${\partial_t N < 0}$ and~${\partial_t   E > 0}$ in
accordance with the discussion in Sec.~\ref{sec:kinetic}. 

Moreover, integrating the left- and right-hand sides of the profile
equation~\eqref{eq:4} over~$\omega_s$ and using the scattering
integral~\eqref{eq:2}, we relate rescaled  charges
to fluxes, 
\begin{align}
\label{eq:N_J}
& k_NN_s= t_{\mathrm{rel}}\,{\cal J}_{N,s}(0) + \int\limits_{0}^{\infty}
J_{\mathrm{ext},s}\, d\omega_s  \,, \\ 
\label{eq:E_J}
& k_EE_{s}= \int\limits_{0}^{\infty} \omega_s J_{\mathrm{ext},s}\, d\omega_s \,,
\end{align}
where we assumed fast falloff of the profile as~${\omega_s=\infty}$
and imposed boundary condition~\eqref{eq:1} at~${\omega_s
=0}$: ${\cal J}_{E,s}(\infty)  ={\cal J}_{N,s}(\infty) = {\cal J}_{E,s}(0) = 0$. 
 
Relations~\eqref{eq:N_J}, \eqref{eq:E_J}  inform us that all energy
deposited by the external source remains in the system whereas the
particle number can leak through the phase-space
boundary~${\omega_s=0}$ or come from beyond it.  In
particular,~$J_{\mathrm{ext},s}$ can be positive-definite (i.e., heat
all parts of the gas) only at~${D>5/2}$. Besides,~$D<3$ and
positive~$J_{\mathrm{ext},s}$ guarantee outward-directed
condensation flux~${\cal J}_{N,s}(0) <0$.  

\begin{figure}
  \centerline{\includegraphics{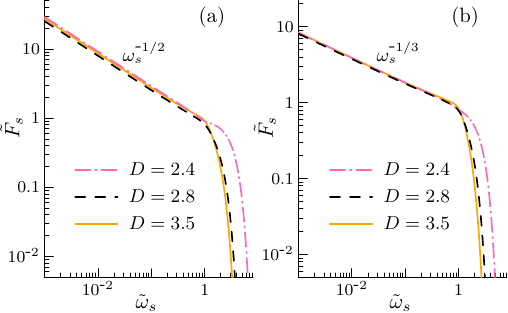}}
  \caption{Self-similar profiles at~$J_{\mathrm{ext}, s}\ne 0$ with
    low-energy tails (a)~${F_s \propto \omega_s^{-1/2}}$ and (b)~${F_s \propto  
      \omega_s^{-1/3}}$. We use dimensionless
    units~(\ref{eq:12}),  boundary condition~\eqref{eq:1}, and  energy
    source~(\ref{eq:28}).} 
  \label{fig:Fs_ws_J}
\end{figure}

Figure~\ref{fig:Fs_ws_J}(a) displays self-similar profiles
at~${J_{\mathrm{ext},s} \ne 0}$ with  different~$D$; we used
boundary condition~\eqref{eq:1} and 
the source
\begin{equation}
    \label{eq:28}
    J_{\mathrm{ext}, s}(\omega_s) =
    \frac{J_0}{\mathrm{ch}^{2}  
    \left[\omega_s/\sigma - \omega_1 \right]}
  \end{equation}
obtained by self-similar rescaling~\eqref{eq:23} of
Eq.~(\ref{eq:19}). As before,~${\sigma = 0.4\,\omega_0}$
and~${\omega_1 = 5}$, but now we fix condensation flux~${{\cal
    J}_{N, s}(0) = - N_0/(2t_{\mathrm{rel}})}$ of all solutions by
tuning the source amplitude~$J_0$.  Despite  seemingly identical
tail-falloff structure, the profiles in Fig.~\ref{fig:Fs_ws_J}(a)
describe different physics, cf.\ Fig.~\ref{fig:ENranges}. At~${D =
  2.4}$ the negative source consumes particles and energy. The
value of~$E$ grows at~${D = 2.8}$ and~$3.5$, but these two cases
differ by the sign of~${\partial_t N}$~--- negative and positive,
respectively.

So far, we studied only profiles with thermal Ray\-leigh–Je\-ans tails at
low energies and exponential cutoffs at high~$\omega_s$. One wonders
about other possibilities. 

On the one hand, power-law high-energy behavior ${F_s \propto
  \omega_s^{\gamma}}$ is totally possible, just not interesting. Indeed,
our scattering integral is UV convergent only at~${\gamma < 0}$, see
Eq.~(\ref{eq:StFcascade}). In this healthy case $\mathrm{St}\, F_s \propto
\omega_s^{3\gamma-1}$ is subdominant to~${F_s   \propto
  \omega_s^{\gamma}}$ as~$\omega_s\to +\infty$~--- hence, the 
profile~$F_s$ is determined by the left-hand side of the kinetic
equation~\eqref{eq:4}. We obtain~${\gamma = 1/(2-D)}$ or 
\begin{equation}
  \label{eq:35}
  F_s \to F_{\infty}\, \omega_s^{1/(2-D)} \qquad \mbox{as} \quad
  \omega_s \to +\infty\,,
\end{equation}
where~${D > 2}$ gives~$\gamma<0$. This is a static ensemble of spectator 
particles which do not participate in gravitational 
scattering: $\mathrm{St}\, F_s$ is negligible and
the original distribution equals~${F   \to F_{\infty} \,\omega^{1/(2-D)}}$,
cf.\ Eqs.~(\ref{eq:an}), (\ref{eq:ab}). We do not consider
such a situation in what follows.

Assuming exponential falloff at high~$\omega_s$, in
  App.~\ref{sec:exponential_infty} we extract 
asym\-pto\-tics of well-localized self-similar profiles,
\begin{equation}
  \label{eq:30}
  F_s \to F_{\infty} \, \omega_s^{\frac{4-D}{2D-4}} \,
  \mathrm{e}^{- c_\infty\omega_s^{5/2}} \quad \mbox{as} \quad
  \omega_s \to \infty\,, 
\end{equation}
where~$F_\infty$, ${c_\infty \equiv 3 N_0^2 (D-2)/(20 D\omega_0^3 G_0)}$,
and $G_0 = \int_{0}^{\infty} d\omega_s \sqrt{\omega_s}
F_s^2$ are constants. This universal behavior is present in all our
  numerical profiles, see the dashed line in Fig.~\ref{fig:Fs_D2.5}.
It requires~$c_{\infty}>0$,~i.e.
\begin{equation}
  \label{eq:27}
  \mbox{exponential falloff}: \quad D>2\,.
\end{equation}
Also, Eq.~(\ref{eq:30}) guarantees UV convergence of the scattering integral
and zero fluxes at~$\omega = +\infty$.

On the other hand, low-energy asymptotics is far more
tricky, since long-range gravitational scattering becomes essential
as~$\omega_s \to 0$. We go around this difficulty by considering only
power-law behaviors~$F_{s} \approx F_0\, \omega_s^{\gamma}$ in
the region~${0 \leq\omega_s \ll   \omega_0}$ and thus returning to the
Kolmogorov-Zakharov setup from Fig.~\ref{fig:cascade}, albeit with
time-dependent~$F_0$. Despite differences with
Sec.~\ref{sec:absence-power-law}, the only two solutions are again 
\begin{equation} 
  \label{eq:29}
  F_s\propto \omega_s^{-1/2}\quad \mbox{or} \quad 
  \omega_s^{-1/3} \qquad \mbox{as} \quad \omega_s \to 0\,.
\end{equation}
Indeed, particle flux equals the primitive of the scattering
integral~(\ref{eq:StFcascade}):~${{\cal J}_{N,s}
  \propto\omega_{s}^{3\gamma}}$. In
nontrivial case it remains nonzero as~$\omega_s \to 0$,
hence,~${\gamma\leq 0}$. For these powers, $\mathrm{St}\,
F_s\propto \omega_s^{3\gamma-1}$  dominates in the profile equation at 
low~$\omega_s$. Thus, kinetic solutions appear when it 
vanishes by itself, i.e.\ in the case~(\ref{eq:29}), see
Eq.~(\ref{eq:StFcascade}). 

Surprisingly, this result does not mean that the low-energy parts of
self-similar solutions are stationary or trivial. Indeed, the
transformation~(\ref{eq:an}) makes both asymptotics~(\ref{eq:29}) 
time-dependent. Besides, in App.~\ref{sec:asympt-as-omeg} we solve
the profile equation via consistent expansion in powers
of~$\omega_s$. In the thermal~$\omega_s^{-1/2}$  case we get,
\begin{equation}
  \label{eq:32}
  F_s \to  F_0 \, \omega_s^{-1/2} + F_3\, \omega_s + O(\omega_s^{3/2})
  \quad \mbox{as} \quad  \omega_s\to 0
\end{equation}
and fluxes~${{\cal J}_{E,s}(0) = 0}$,~${{\cal
    J}_{N,s}(0) = - 15 W_0\, F_0^2   F_3 / 
  4}$. We see that~$\omega_s$ corrections generate nonzero particle
flux through the phase-space boundary. All solutions in
Fig.~\ref{fig:Fs_ws_J}(a) satisfy Eq.~(\ref{eq:32}).

In the~$\omega_s^{-1/3}$ case power-law expansion of
App.~\ref{sec:asympt-as-omeg} yields,
\begin{equation}
  \label{eq:31}
  F_s \to F_0\, \omega_s^{-1/3} + F_1\, \omega_s^{2/3} +
  O(\omega_s^{4/3}) \quad \mbox{as} \quad  \omega_s\to 0\,.
\end{equation}
We get finite particle flux~${{\cal J}_{N,s}(0)= -729 W_0 F_0^2
  F_1/220}$ and negative energy flux ${{\cal J}_{E,s}(0) \;
  = \;-9 W_0  F_0^3 / 40}$,  cf.\ Eq.~(\ref{eq:20}). This means that
solutions (\ref{eq:31}) break the   boundary condition~\eqref{eq:1} but comply
with more general finite-flux requirements that will be formulated in
Sec.~\ref{sec:universal-limit}. Treating Eq.~(\ref{eq:31}) as a
boundary condition, we successfully compute the~$\omega^{-1/3}$ 
numerical profiles with the source~(\ref{eq:28}), see
Fig.~\ref{fig:Fs_ws_J}(b). All of them have~${{\cal J}_{E,s}(0) < 0}$.

It is worth noting that the~$\omega_s^{-1/3}$ profiles stem from a
smaller family as compared to the solutions
with~$\omega_s^{-1/2}$ tails. In the next subsection we will 
generalize the latter to have an arbitrary energy flux~${\cal
  J}_{E,s}$~at~${\omega_s=0}$. The same parameter is strictly fixed
in the~$\omega^{-1/3}$ solutions making them fine-tuned.

Recall that the thermal $\omega_s^{-1/2}$ asymptotics of
self-similar solutions can be strongly deformed by finite-width
sponges, and this property persists at~$J_{\mathrm{ext}} \ne 0$.
For example, the chain-point graph in Fig.~\ref{fig_attract_kinJ}(b)
  solves the same profile equation with~$D=2.8$ as dashed
  curve in Fig.~\ref{fig:Fs_ws_J}(a) but satisfies
  boundary condition~\eqref{eq:25} with
  finite~$\omega_{\mathrm{IR},s}=10^{-3}\, \omega_0$.
As a result, it grows at  low~$\omega_s$ unlike the profile in
Fig.~\ref{fig:Fs_ws_J}(a). 

We also stress that all our profiles with~$\omega^{-1/2}$ tails
are stable kinetic attractors. In particular, the simulation in
Fig.~\ref{fig_attract_kinJ}(b) approaches the one with~$D=2.8$.
We computed kinetic evolutions with~${D=2.4}$ and~$3.5$  time
  dependences of self-similar source~\eqref{eq:19}
   and obtained the same result: time-dependent
  distributions invariably approached scaling profiles
  with respective~$D$.

In contrast, we failed to observe low-energy behavior~${F \propto
  \omega^{-1/3}}$ in time-dependent numerical simulations, even if
with distortions. We tried starting evolutions from the
profiles in Fig.~\ref{fig:Fs_ws_J}(b), but they shortly decayed via
${\mbox{low-}\omega}$ instabilities. We even failed to promote
the~$\omega_s\to 0$ boundary condition~(\ref{eq:31})  to 
realistically-looking setup with finite-width sources and sponges: any
modification caused the~$F_s \propto\omega_s^{-1/3}$ 
tails to disappear. This strongly indicates instability of
self-similar profiles with~$\omega_s^{-1/3}$ behavior at low energies. 

\subsection{Universal limit}
\label{sec:universal-limit}

Our effective approach has one drawback: scaling profiles are dictated
by hand-made sources~$J_{\mathrm{ext},  s}(\omega_s)$ which
render them non-specific. In this Section we consider universal
limit of narrow source sitting at the phase-space boundary~${\omega_s
  = 0}$.

\begin{figure}
  \centerline{
    \unitlength=1mm
    \begin{picture}(84, 50)
      \put(0,0){\includegraphics{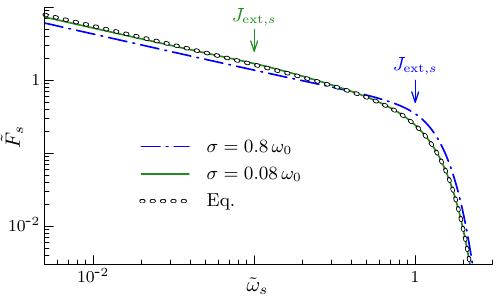}}
      \put(41.5,15){\eqref{eq:33}}
    \end{picture}}
  \caption{Two self-similar profiles (dash-dotted to solid line) with
    sources~\eqref{eq:28} of shrinking width~$\sigma$ and decreasing
    central position~${\omega_s = \omega_1 \sigma}$  (arrows above the
    graphs). We use $D = 2.8$, $\omega_1 = 5/2$, and boundary
      condition~\eqref{eq:1}. Source amplitudes~$J_0$ are selected to
    fix the rescaled particle number~${N_{s} =
      N_0}$ of solutions. Chain points show universal profile
    with~${J_{\mathrm{ext},s}=0}$, boundary condition~\eqref{eq:33},
    ${{\cal J}_{E,s}(0)\approx 0.157\, \omega_0N_0/t_{\mathrm{rel}}}$,
    and~${N_s=N_0}$.}
  \label{fig:to_universality}
\end{figure}

Namely, let us gradually decrease the width ${\sigma\to 0}$ of
  the source~\eqref{eq:28} at ${\omega_1=5/2}$. We keep
fixed the rescaled particle number~${N_{s} = N_0}$ of solutions
by tuning the source amplitude~$J_0$. A succession of two
such profiles with~${D=2.8}$ is shown in
Fig.~\ref{fig:to_universality}. They almost coincide and hence have a
well-definite limit as~${\sigma\to 0}$~--- a configuration
with~$\delta$-source sitting precisely at the
point~${\omega_s=0}$.
  
The latter limiting solution satisfies the profile
equation~\eqref{eq:4} with~$J_{\mathrm{ext},s}=0$ but has nonzero
fluxes of particles and energy coming from the lowest-energy
  region. A suitable boundary condition in this case is 
\begin{equation}
  \label{eq:33}
  {\cal J}_{E,s} = {\cal J}_{E,s}^{\mathrm{fixed}} \quad \mbox{and}
  \quad |{\cal J}_{N,s}| < \infty  \quad
  \mbox{at} \quad \omega_s=0\,;
\end{equation}
it replaces Eq.~\eqref{eq:1}. Imposing Eq.~(\ref{eq:33}), we
compute the numerical profile with ${J_{\mathrm{ext},s}=0}$ and
${D=2.8}$ tuning ${{\cal J}_{E,s}^{\mathrm{fixed}}\approx 0.157\,
  \omega_0N_0/t_{\mathrm{rel}}}$ to set~${N_s = N_0}$, see the chain
points in Fig.~\ref{fig:to_universality}. The latter profile is
close to the ${\mbox{smallest-}\sigma}$ graph, indeed. We remind that
the actual~${\omega=0}$ fluxes of this self-similar solution are
time-dependent, 
\[
  {{\cal J}_{N} =
    \alpha^{3}(\tau) {\cal J}_{N,s}(0)} \quad
\mbox{and} \quad
     {{\cal J}_{E}  = \alpha^{3}(\tau)  {\cal J}_{E,s}(0) /
       \beta(\tau)},
\]
for fixed~${\cal J}_{E,s}(0)$ and~${\cal J}_{N,s}(0)$, see Eqs.~(\ref{eq:an}), (\ref{eq:14}).

\begin{figure}
  \centerline{\includegraphics{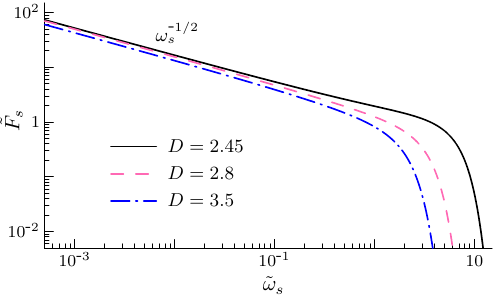}}
  \caption{Universal self-similar profiles with arbitrary~$D$
      and ${|{\cal J}_{E,s}(0)| = 2\omega_0
        N_0/t_{\mathrm{rel}}}$. They have~$J_{\mathrm{ext},s}=0$ and
      satisfy the boundary conditions~\eqref{eq:30},~\eqref{eq:33}}.
    \label{fig:Fw_ws_universal}
\end{figure}

We checked that self-similar solutions with  sources at extremely
low~$\omega_s$ remain attractors of kinetic evolution. To this end we
started time-dependent simulation from the one with~${D=2.8}$,
the cutoff~\eqref{eq:25} at ${\omega_{\mathrm{IR},s} = 2\cdot10^{-3}\,
  \omega_0}$, and the source~\eqref{eq:28} of strength~${J_0 = 1267
  N_0/\omega_0}$ at low~${\omega_s \sim \omega_1\sigma = 10^{-2}\,
  \omega_0}$. We evolved the source according to
Eq.~\eqref{eq:19} with~${D=2.8}$ and~${\tau_i = -1}$ keeping it
immersed deep inside the~$\omega^{-1/2}$ tail of the profile.
Nevertheless, the time-dependent solution remained stable during the 
simulation period of~$95$ relaxation times and its final distribution
reproduced  (via rescaling) the profile we started with.

One naturally parametrizes universal self-similar profiles by~$D$
and~${\cal J}_{E,s}^{\mathrm{fixed}}$, see App.~\ref{sec:self-simil-prof} for
proper count of free parameters. However, residual scale
symmetry~${F_s \to F_s(b^2\omega_s)/b}$ of the sourceless profile
equation leaves~${\cal J}_{E,s}^{\mathrm{fixed}}$ the role of normalization
constant. This turns Eq.~\eqref{eq:4} with~${J_{\mathrm{ext},s}=0}$
and boundary conditions~\eqref{eq:30}, \eqref{eq:33} into a nonlinear
Sturm-Liouville problem for continuous ``eigenvalues''~$D\geq 2$ and
``eigenfunctions''~$F_s(\omega_s)$.  Figure \ref{fig:Fw_ws_universal}
shows universal profiles at different~$D$.

It is worth noting that the conservation laws~\eqref{eq:N_J},
\eqref{eq:E_J} simplify at~${J_{\mathrm{ext},s}=0}$. Namely,
integrating all terms of Eq.~\eqref{eq:4} over~$\omega_s$, using
the scattering integral~\eqref{eq:2} and boundary condition
(\ref{eq:33}), we get,
\begin{equation}
\label{eq:Csol-1/2}
k_NN_s= t_{\mathrm{rel}}\, {\cal J}_{N,s}(0)   \quad \mbox{and} \quad
k_EE_s= t_{\mathrm{rel}}\, {\cal J}_{E,s}(0)\,.
\end{equation}
These identities relate the signs of~$k_N$ and~$k_E$ in
Eq.~\eqref{eq:k} and Fig.~\ref{fig:ENranges} to directions of
particle and energy fluxes through the boundary~${\omega_s = 0}$. 
For example, the profiles with~${D=2.45}$ and~$2.8$,~$3.5$ in
Fig.~\ref{fig:Fw_ws_universal} have negative (outward-directed) and
positive (incoming)~${{\cal J}_{E,s}(0) \propto \partial_t E}$, respectively.

Of course,  boundary conditions~(\ref{eq:33}) change infrared
asymptotics of self-similar profiles: now\footnote{We do not consider
universal solutions with unstable~$\omega^{-1/3}$ tails.} 
\begin{equation}
  \label{eq:34}
  F_s =  F_0 \, \omega_s^{-1/2} + F_1 - \frac{F_1^2}{2F_0} \,
  \omega_s^{1/2} + F_3\, \omega_s + O(\omega_s^{3/2})\,,
\end{equation}
at $\omega_s\to 0$, cf.\ Eq.~\eqref{eq:32} and see
App.~\ref{sec:asympt-as-omeg} for derivation.  The respective fluxes
${{\cal J}_{E,s}(0) =  -3 W_0 F_0^2 F_1/2}$ and ${{\cal
      J}_{N,s}(0) =  W_0(7F_1^3 -15  F_0^2 F_3)/4}$ have arbitrary
signs. On the other hand, high-energy behavior of universal profiles
is still determined by Eq.~\eqref{eq:30}.   


\section{Adiabatic self-similarity}
\label{sec:compare}
\subsection{Numerical example}
\label{sec:mot}

Now, we move to more realistic situation of broken scale
symmetry~\eqref{eq:11}. This can be done by  enabling sources with
generic time dependence, say, an arbitrary power-law
\begin{equation}
  \label{eq:37}
  J_{\mathrm{ext}}(\tau,\, \omega) = \frac{J_0}{t_{\mathrm{rel}}}\,
  \, \frac{\tau^{-3/4} \, \theta(\tau-1)}{\mathrm{ch}^{2}(\omega/\sigma -  \omega_1)}\,,
\end{equation}
voluntarily turning on at~${\tau\geq 1}$. Here~$\tau \equiv
t/t_{\mathrm{rel}}$,  whereas~$J_0$, $\sigma$, $\omega_{1}$ are
constants; cf.\ Eq.~\eqref{eq:19}. Thanks to symmetry
breaking, the Ansatz~\eqref{eq:an}, \eqref{eq:ab} does
not pass the kinetic equation. But we will see that 
the evolution stays remarkably close to self-similar profiles
nevertheless.

\begin{figure}
  \centerline{\includegraphics{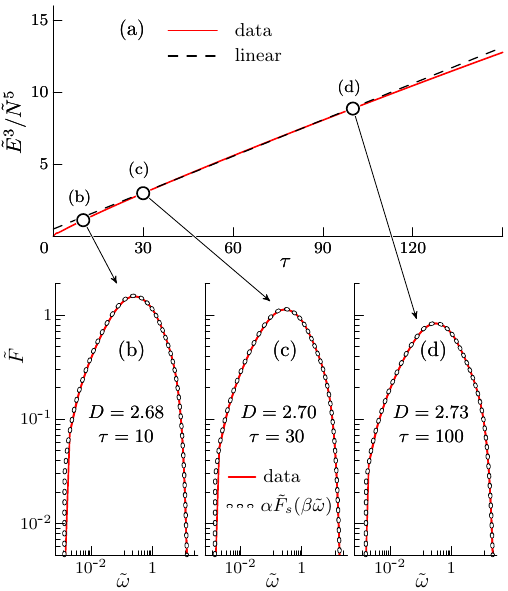}}
  \caption{Numerical solution of time-dependent kinetic
    equation~\eqref{eq:13} with scale-breaking  source (\ref{eq:37}),
    sponge~\eqref{eq:17}, and Gaussian initial
    distribution~\eqref{eq:16} (solid lines). We use dimensionless 
    units~\eqref{eq:12},  ${\tilde{N} \equiv N/N_0}$,
    ${\tilde{E}\equiv E/(2\omega_0 N_0)}$, and parameters~${J_0  =
      5N_0/(8\omega_0)}$, ${\sigma = 0.8\, \omega_0}$,  ${\omega_1 =
      5/2}$. (a)~Combination~$E^3(\tau)/N^5(\tau)$ computed on
    the solution (solid line) and a linear function of~$\tau$
    (dashed). (b),~(c),~(d)~Numerical distribution function~$F(\tau,\,  
    \omega)$ at fixed~$\tau$ (solid lines) compared to rescaled
    self-similar profiles (chain points). Scaling
    dimensions~${D(\tau)}$ of the latter are indicated on the panels.}
    \label{fig:com1}
\end{figure}  

We take the Gaussian initial data~\eqref{eq:16} with~${c_N =43.4}$, add the
  sponge~\eqref{eq:17} and scale-breaking source~(\ref{eq:37}) to the
  kinetic equation~\eqref{eq:13}, and solve it numerically. One observes two
    eye-catching patterns in the result of this time-dependent
simulation, see Fig.~\ref{fig:com1}. First, the combination~$E^3(\tau)/N^5(\tau)$ 
of the solution energy and particle number evolves almost linearly with time,
see the panel~(a). This is the  hallmark of self-similar dynamics: 
Eqs.~\eqref{eq:26} and~\eqref{eq:k} predict~${E^{3}/N^{5} \propto   \tau -
  \tau_i}$. Second, the solution~${F(\tau,\, \omega)}$ itself is
almost indistinguishable at every~$\tau$ from the rescaled
self-similar profiles~${\alpha(\tau) F_s(\beta(\tau) \omega)}$, cf.\ solid lines
and chain points in Figs.~(b), (c), and~(d). The latter scaling solutions,
however, have different dimensions:~${D = D(\tau)}$. 

It is clear, what happened. Since self-similar solutions are
kinetic attractors, the source (\ref{eq:37}) does not ruin
them completely but enforces adiabatic drift of their parameters
with time. In the next subsection we will learn to describe this regime
of adiabatic (approximate) self-similarity. 

\subsection{Adiabatic approximation}
\label{sec:apprx}

Suppose kinetic evolution is secretly almost self-similar with
time-dependent scaling dimension~${D = D(\tau)}$. We introduce
scaling coefficients  
\begin{align}
  \label{eq:ab_new}
  & \alpha(\tau) = \frac{1}{\sqrt{1-\tau_i}}\exp\left[ - \int_{1}^{\tau}
    \frac{1}{D(\tau')} \frac{d\tau'}{\tau'-\tau_i}\right]\,,
  \\[1ex] \notag
  & \beta(\tau)  =
  \exp\left[\int_{1}^{\tau} \bigg(\frac{2}{D(\tau')}
    -1\bigg)\frac{d\tau'}{\tau'-\tau_i}\right]\,,
\end{align}
which reduce to powers of time~\eqref{eq:15} for
  time-in\-de\-pen\-dent~$D$;~$\tau_i$ is a constant\footnote{More accurate
approach may use non-stationary~${\tau_i = \tau_i(\tau)}$ at the 
price of inventing additional equation for this parameter.}
and we made the integrals run from~${\tau=1}$.   

The first step towards adiabatic self-similarity is to regard
the self-similar Ansatz~(\ref{eq:an}) as a change of
variables
\begin{equation}
\label{eq:FsJs}
F(\tau, \, \omega) = \alpha(\tau)F_s(\tau_s,\, \omega_s) \,,
\end{equation}
where~${\tau_s  \equiv\ln(\tau-\tau_i)}$ is the new time
and~${\omega_s \equiv \beta(\tau)\,\omega}$. In new terms, kinetic
equation~\eqref{eq:13} looks like   
\begin{multline}
  \label{eq:Fs-s}
  \partial_{\tau_s}F_s
  -F_s/D + \left(2/D-1\right)\omega_s\partial_{\omega_s}F_s 
  \\[.7ex] =t_{\mathrm{rel}}\, \mathrm{St}\, F_s+J_{\mathrm{ext},s}(\tau_s,\, \omega_s)\,,
\end{multline}
where~$J_{\mathrm{ext},s}$ is still given by Eq.~\eqref{eq:23} but now
depends on~$\tau_s$. We mimic static absorbing sponge at~${\omega
  \lesssim \omega_{\mathrm{IR}}}$ by boundary condition 
\begin{equation}
  \label{eq:40}
  F_s=0 \qquad\mbox{at} \qquad
  \omega_s \leq \omega_{\mathrm{IR}}/\beta(\tau)\,,
\end{equation}
cf.\ Eq.~\eqref{eq:25}.

We did not make any significant approximations yet: up to
  details\footnote{Strictly speaking, Eq.~\eqref{eq:40} is valid
    only approximately. Note, however, that Eq.~(\ref{eq:Fs-s}) can be 
    generalized to the full case of nonzero sponge~$\mu(\omega)$ acting
    at~${\omega \leq \omega_{\mathrm{IR}}}$,
    cf.\ Eq.~\eqref{eq:13}. We use simpler Eq.~\eqref{eq:40} only
    because it is very precise for strong narrow sponges we
    consider.}, the boundary value problem~(\ref{eq:Fs-s}),
  (\ref{eq:40}) is equivalent to the original kinetic equation. But if
  adiabatic self-similarity reigns 
and parameters of the transformation~(\ref{eq:FsJs}) are
chosen correctly, 
the rescaled distribution~$F_s(\tau_s,\, \omega_s)$ slowly varies
  with~$\tau_s$. Hence, the first term~$\partial_{\tau_s} F_s$ in 
Eq.~(\ref{eq:Fs-s}) can be ignored or treated as a perturbation.

Strictly speaking, adiabatic approximation of this kind requires
suppressed scale-breaking effects, i.e.\ slow drift of the rescaled  
source~$J_{\mathrm{ext},s}(\tau_s, \,\omega_s)$, see the
condition~\eqref{eq:41}. But in practice adiabatic results are
qualitatively correct even in generic cases.

In the leading adiabatic order, we ignore the time derivative
of~$F_s$ in Eq.~(\ref{eq:Fs-s}) and arrive at the profile
equation~\eqref{eq:4} with nonstationary
source~$J_{\mathrm{ext},s}$. Now, the time enters the problem as a
parameter. This allows us to solve at every~$\tau$
Eqs.~\eqref{eq:4}, (\ref{eq:40}) for slowly-drifting
profile~${F_s = F_s^{(0)}(\tau_s,   \,\omega_s)}$, and then, by means of
reverse self-similar transformation  (\ref{eq:FsJs}), obtain an
approximate solution of the original kinetic equation.

It is worth noting that higher-order corrections can be consistently
incorporated in the adiabatic approach. Namely, subtract the time
derivative of  the leading-order profile from the
source,~${J_{\mathrm{ext},s} \to J_{\mathrm{ext},s}   -
  \partial_{\tau_s} F_s^{(0)}}$, then solve the profile
equation~\eqref{eq:4} with such new source in the right-hand
side. The result it more accurate second-order  
profile~$F_s^{(1)}$. Further subtractions-solvings provide
higher-order solutions~$F_s^{(n)}$ which at~${n\to +\infty}$ converge
to exact kinetic evolution, cf.\ Eqs.~\eqref{eq:Fs-s}
  and~\eqref{eq:4}. We will disregard corrections in what  
follows: the profile~$F_s^{(0)}$ will be  sufficient.

Now, to the important question. Equation~\eqref{eq:Fs-s} is valid for
any~$D(\tau)$ and~$\tau_i$, but our adiabatic approximation is
not: correct choice of these parameters is needed for good
precision. We select them on case-by-case basis using conservation
laws.

Consider, e.g., the setup in Sec.~\ref{sec:mot}: a predefined
source~$J_{\mathrm{ext}}(\tau,\, \omega)$ heats the gas condensing 
into a sponge. Evolution of the gas energy~$E(\tau)$
is controlled exclusively by the source because the sponge eats
particles with~$\omega \approx 0$, cf.\ 
Eq.~\eqref{eq:17}. Integrating the left- 
and right-hand sides of the kinetic equation~\eqref{eq:13}
over~$\omega$ and ignoring the sponge,  we get, 
\begin{equation}
  \label{eq:38}
  \partial_\tau E   \approx
  t_{\mathrm{rel}}\int_{0}^{\infty}  J_{\mathrm{ext}}(\tau,\,
  \omega')\,\omega'  d\omega' \,,
\end{equation}
where the the right-hand side and hence~$E(\tau)$ can be
  explicitly computed, e.g.,  for the  source~\eqref{eq:37}
of Sec.~\ref{sec:mot}.

It is clear that correct parameter~$D(\tau)$ should reconcile evolution
of energy in Eq.~(\ref{eq:38}) with self-similar law~\eqref{eq:21}. We
define it using the relation
\begin{equation}
\label{eq:fixD}
\frac{\partial_\tau E}{E}  \equiv \frac{k_E(\tau)}{\tau-\tau_i} \,,
\quad \text{where}\quad k_E=2 - \frac{5}{D(\tau)}\,,
\end{equation}
giving explicit function ${D(\tau) = 5/(2-\partial_{\tau_s} \ln E)}$.

In truth, Eq.~\eqref{eq:fixD} automatically sets the
energy~$E^{(0)}(\tau)$ of the leading-order adiabatic solution 
equal  to the exact energy~$E(\tau)$. Indeed, the leading-order profile 
equation~\eqref{eq:4} for~$F_s^{(0)}$ has its own
conservation law~\eqref{eq:E_J}, 
\begin{equation}
\label{eq:fixk_e}
k_EE_s^{(0)} \approx
\int\limits_{0}^{\infty}\omega_s'd\omega_s'\,  J_{\mathrm{ext},s}(\tau,\,\omega_s')
\approx \frac{\beta}{\alpha^3} \, \partial_\tau E(\tau)\,,
\end{equation}
where in the second equality we performed inverse
transformation~\eqref{eq:23} of the source and exploited
Eq.~(\ref{eq:38}). Identity~\eqref{eq:fixk_e},
definition~\eqref{eq:fixD} of~$D$, 
and self-similar transformation~$E_s^{(0)} \equiv \beta^{2}
E^{(0)}/\alpha$ imply~$E^{(0)}(\tau) = E(\tau)$, indeed. 

We fix the last self-similar parameter~--- reference time~$\tau_i$~---
by using the Cauchy data, namely, by equating the
masses~${M^{(0)}(\tau_2) \equiv M(\tau_2)}$ of approximate and exact
solutions at some~${\tau = \tau_2}$. The latter time moment will
be taken large enough for the regime of adiabatic self-similarity to
set in.

\begin{figure}
  \centerline{
    \unitlength=1mm
    \begin{picture}(86,80)
      \put(0,0){\includegraphics{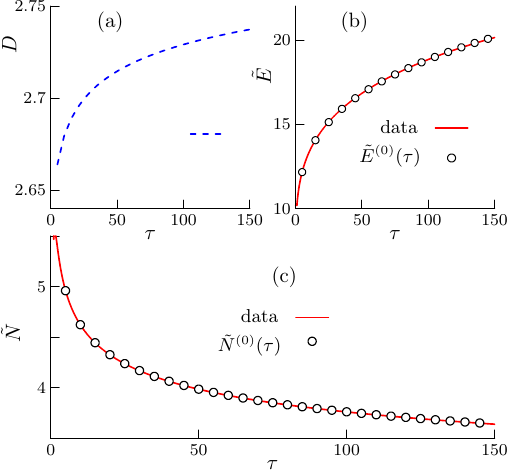}}
      \put(19,56.4){Eq.$\,$\eqref{eq:fixD}}
    \end{picture}}
  \caption{Numerical solution of the kinetic equation~\eqref{eq:13}
    with scale-breaking  source (\ref{eq:37}) and 
    sponge~\eqref{eq:17} (solid lines) compared to the leading-order
    adiabatic result (circles); see also
    Fig.~\ref{fig:com1}. (a)~Scaling dimension~$D(\tau)$ in 
    Eq.~(\ref{eq:fixD}) (dashed line). (b),~(c)~Evolution of exact and
    approximate energies~${E(\tau) \approx E^{(0)}(\tau)}$ and
    multiplicities~${N(\tau) \approx  N^{(0)}(\tau)}$.} 
 \label{fig_M}
\end{figure}

The above receipt for finding~$D(\tau)$ and~$\tau_i$ finishes
formulation of our adiabatically self-similar approach. Let us apply
it to the specific source~\eqref{eq:37}. 
Definition~\eqref{eq:fixD} provides~$D(\tau)$ changing
in a relatively wide interval ${2.65 \lesssim D  \lesssim 2.75}$, see
Fig.~\ref{fig_M}(a) and cf.\  Fig.~\ref{fig:ENranges}. Next, we solve
the profile equation~\eqref{eq:4} with given~$D(\tau)$, 
cutoff~\eqref{eq:40}, and transformed
source~\eqref{eq:37},~\eqref{eq:23} at different~$\tau$.  The final
self-similar parameter~${\tau_i\approx-0.0066}$  is fixed by
equating the approximate and exact masses,\footnote{In practice,
one solves this equation by 
bisection of~$\tau_i$.}~${M^{(0)}(\tau_2) = M(\tau_2)}$,
at~${\tau_2=5}$. We arrive at the  leading-order 
solution ${F^{(0)}(\tau,\,  \omega)}$ which has~$E^{(0)}(\tau) = E(\tau)$
thanks to the choice of~$D(\tau)$, see Fig.~\ref{fig_M}(b). But what
is far less trivial, our adiabatic distribution function~${F^{(0)}
  \approx F}$ and total mass~${M^{(0)} \approx M}$ are also
  close to the simulation results, see the lines vs.\ chain points in 
Figs.~\ref{fig:com1}(b), (c), (d) and the line vs.\ circles 
in Fig.~\ref{fig_M}(c). Note that the accuracy of our adiabatic
solution is even better than the estimate~$\partial_{\tau_s}
\ln D \sim 10^{-2}$ in Eq.~\eqref{eq:41}.

Naturally, the method of adiabatic self-similarity can be
extended beyond the toy model of Sec.~\ref{sec:mot}. Below we 
consider its physically motivated application. 

\subsection{Growth of Bose stars}
\label{sec:growth-bose-stars}

At last, we consider kinetic growth of dark matter Bose stars in the
gas of dark bosons. We have already pointed out in
Sec.~\ref{sec:kinetic}   that this process can  be modeled by kinetic
evolution~\eqref{eq:13} with properly adjusted  source~$J_{\mathrm{ext}}$
and sponge~$\mu$. But acceptable expressions for the latter do not
exist yet, cf.\ Ref.~\cite{Chan:2022bkz, Dmitriev:2023jnu}, so we will rely on heuristic
observation of Ref.~\cite{Dmitriev:2023ipv} that  the gas near the
Bose star evolves in adiabatically self-similar way. We  will see that
this suffices for solving the growth problem.  

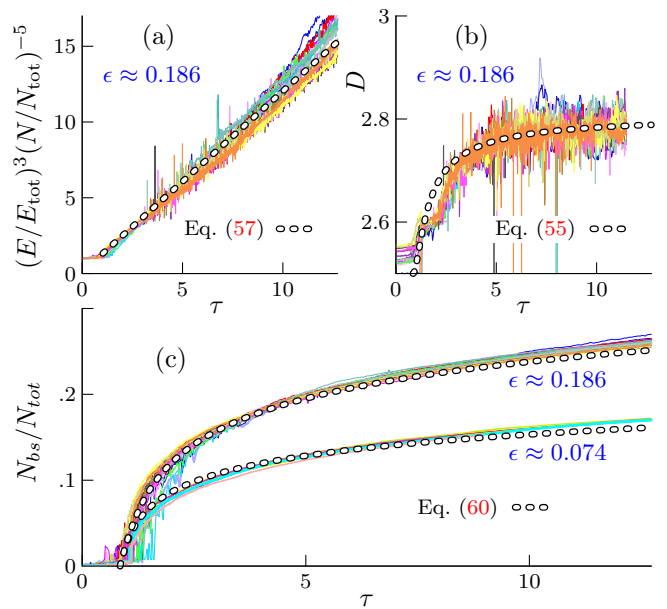
\begin{figure}
  \centerline{\unitlength=1mm
    \setlength{\fboxsep}{1pt}
    \begin{picture}(86,80)
    \put(0,0){\includegraphics{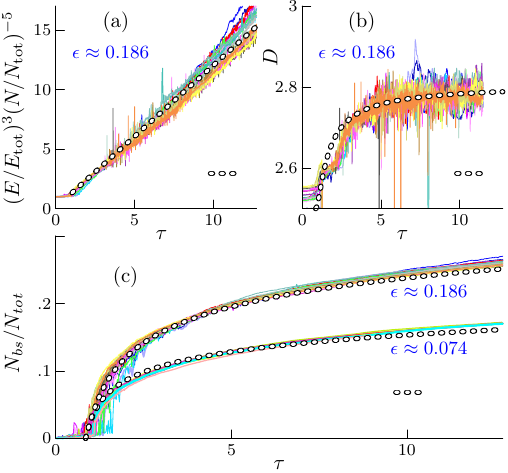}}
    \put(23,49.8){\footnotesize Eq.~\eqref{eq:E3M5}}
    \put(64,49.8){\footnotesize \colorbox{white}{Eq.~\eqref{eq:39}}}
    \put(53.7,12.7){\footnotesize \colorbox{white}{Eq.~\eqref{eq:44}}}
    \end{picture}}
   \caption{Results of 22 (11) microscopic
     Schr\"odinger-Poisson simulations at $\epsilon \approx 0.186$
       ($\epsilon \approx 0.074$)
     of the gas-Bose-star system  in Ref.~\cite{Dmitriev:2023ipv}
     (thin fluctuating lines) compared to predictions of adiabatically
     self-similar approach (chain points). Best-fit heuristic
     parameters of the latter are~${\tau_i \approx -0.085}$ and~${x_e
       \approx 0.043}$ or~$0.021$ at~${\epsilon \approx 0.074}$ 
     or~$0.186$. Normalization constants~$E_{\mathrm{tot}}$
       and~$N_{\mathrm{tot}}$
       represent total energy and multiplicity of the system.
     (a)~Ratio~${E^3/M^{5}}$ vs self-similar
     law~\eqref{eq:E3M5}. (b)~Scaling dimension~$D(\tau)$ estimated
     from Eqs.~\eqref{eq:39},~\eqref{eq:k}:~${D \approx
       (5-3R)/(2-R)}$,  where ${R \equiv \partial_\tau\ln
       E/\partial_\tau\ln N}$. Prior to differentiating, we suppressed
     fluctuations of exact~$E(\tau)$ and~$N(\tau)$ by averaging them
     over moving window~$\Delta \tau = 2.5$. (b)~Bose star
     multiplicity~$N_{bs}(\tau)$ at two values of~$\epsilon$.}
  \label{fig:Mbs}
\end{figure}

This time, the definition~\eqref{eq:fixD} of~$D(\tau)$ is
impractical because evolution of the gas energy is not
known in advance. But we have another good option. Indeed, adiabatic
self-similarity forces~$E(\tau)$ and~$N(\tau)$ to evolve as
powers~\eqref{eq:26} of time
 \begin{equation}
   \label{eq:39}
   \frac{\partial_\tau N}{N} \approx \frac{k_N(\tau)}{\tau - \tau_i} 
   \qquad \mbox{and} \qquad
   \frac{\partial_\tau E}{E} \approx \frac{k_E(\tau)}{\tau - \tau_i}
 \end{equation}
 with related dimensions~$3 k_E - 5k_N \approx 1$ in
 Eq.~\eqref{eq:k}. Let us make this last relation exact in the
 leading order  by adjusting~$D(\tau)$:
 \begin{equation}
   \label{eq:3dlE-5DlM}
   \quad 3\,\frac{\partial_\tau E^{(0)}}{E^{(0)}}-5\,\frac{\partial_\tau
     N^{(0)}}{N^{(0)}}  = \frac{1}{(\tau-\tau_i)}\,.
 \end{equation}
Equality~(\ref{eq:3dlE-5DlM})  immediately  integrates into an
approximate equation of state for a self-similar  bosonic gas, 
 \begin{equation}
   \label{eq:E3M5}
   \frac{\left[E^{(0)}(\tau)\right]^3}{\left[M^{(0)}(\tau)\right]^5} =
   \frac{(\tau-\tau_i)}{\tau_*} \, \frac{E_{\mathrm{tot}}^3}{N_{\mathrm{tot}}^5}\,,
 \end{equation}
 where~$\tau_*$ is an integration constant
 and~$E_{\mathrm{tot}}$,~$N_{\mathrm{tot}}$ are the reference
 ``total'' values to be  specified later. In Fig.~\ref{fig:Mbs}(a)
 we demonstrate that relation~(\ref{eq:E3M5}) (chain points) correctly
 reproduces microscopic (Schr\"odinger-Poisson) simulations of  the
 gas-Bose-star system in  Ref.~\cite{Dmitriev:2023ipv} (fluctuating lines).  

 It is inadvisable to proceed with computation of self-similar
 profiles\footnote{We obtained them in Ref.~\cite{Dmitriev:2023ipv}
 regardless. For properly chosen~$J_{\mathrm{ext}}$ and~$\mu$ they
 are close to the simulation results.} as we did in the previous
 Section. That would force us to specify the source and 
the sponge thus tainting results with functional freedom. Instead, 
we solve   Eq.~\eqref{eq:E3M5} together with conservation 
 laws~\cite{Dmitriev:2023ipv} treating~$\tau_i$  and~$\tau_*$ as 
 heuristic parameters.

Of course, total multiplicity~$N_{\mathrm{tot}}$ and
energy~$E_{\mathrm{tot}}$ are conserved,
\begin{equation}
   \label{eq:42}
   N_{\mathrm{tot}} = N^{(0)} + N_{bs} + N_{e}\,, \quad
   E_{\mathrm{tot}} = E^{(0)} + E_{bs} + E_e\,, 
 \end{equation}
 in the complete system including the gas with~$N^{(0)}$ particles and
 energy~$E^{(0)}$, Bose star with parameters~$N_{bs}$, $E_{bs}$, and a
 part of the gas with charges~$N_e$,~$E_e$ occupying excited
 levels in the Bose star gravitational well,  see
 Fig.~\ref{fig:condensation_dia}(b). Note that  we have already   
 used parameters~$N_{\mathrm{tot}}$ and~$E_{\mathrm{tot}}$ as normalization 
 constants in Eq.~(\ref{eq:E3M5}). Simulations show~\cite{Dmitriev:2023ipv}
 that~$N_e$ is nearly time-independent and~${E_e \approx
     0}$. Then Eqs.~(\ref{eq:42}) relate gas and Bose star charges.

 Here is another bonus: Bose star itself is a gravitationally
 bound object with definite equation of state,
 \begin{equation}
   \label{eq:43}
   E_{bs} = -\gamma N_{bs}^3 \,, \quad \mbox{where} \quad \gamma
   \approx 0.0542 \, m^5 G^2
 \end{equation}
 is a numerical constant; see,  e.g., Ref.~\cite{Dmitriev:2021utv}.

 Substituting Eqs.~(\ref{eq:42}) and~(\ref{eq:43}) into
 Eq.~(\ref{eq:E3M5}), we arrive at the growth law for the Bose star 
 mass~${M_{bs} \equiv m N_{bs}}$, 
 \begin{equation}
   \label{eq:44}
   \frac{(1 +  x_{bs}^3/\epsilon^2)^3}{(1 - x_e - x_{bs})^5} =
   \frac{\tau - \tau_i}{\tau_*}\,, \;\; \mbox{where} \;\; x_{bs}(\tau) \equiv
   \frac{N_{bs}}{N_{\mathrm{tot}}}\,.
 \end{equation}
It involves heuristic parameters~$\tau_*$,~$\tau_i$, and ${x_e
  \equiv N_e/N_{\mathrm{tot}}}$ along with  known
combination\footnote{Related to invariant~${\Xi \approx 0.0542 \,
  \epsilon^2}$ of  Ref.~\cite{Mocz:2017wlg}.} ${\epsilon^2 =
  E_{\mathrm{tot}}/(\gamma N_{\mathrm{tot}}^3)}$ of total energy and
multiplicity.

In fact, one parameter is set by the initial data. Recall that
Bose star forms in the gas at~$\tau = \tau_{gr}  \equiv 2^{3/2} b/3
\approx 0.85$, where~$b\approx 0.9$, see
  Ref.~\cite{Levkov:2018kau}. Imposing~${x_{bs}(\tau_{gr}) = 0}$, we 
obtain,  
\begin{equation}
  \label{eq:45}
  \tau_* = (\tau_{gr} - \tau_i) (1 - x_e)^5\,.
\end{equation}
The last two heuristic parameters are provided by fits of 
numerical data. In Ref.~\cite{Dmitriev:2023ipv} we arrived at
universal value~${\tau_i \approx -0.1\, \tau_{gr}}$ and small~${x_e \approx
0.043}$ and~$0.021$ at~${\epsilon \approx 0.074}$ and~$0.186$,
respectively. With these parameters, the growth law~(\ref{eq:44})
correctly reproduces simulation results, cf.\ the lines and the chain points
in Fig.~\ref{fig:Mbs}(c).

Note that the analytical approach of this subsection is simple but
limited to conserved quantities and incapable of reproducing full
phase-space distribution. Nevertheless, it  relies on
time-dependent scaling dimension~$D(\tau)$. The latter~--- estimated from
Eqs.~(\ref{eq:39})~--- is shown in Fig.~\ref{fig:Mbs}(b) where the
lines and chain points represent simulation results and analytical growth 
 law~\eqref{eq:44},~\eqref{eq:42}, respectively.


\section{Discussion}
\label{sec:discussion}

In this paper we proposed self-similar kinetics for overpopulated
gas of gravitationally interacting bosons surrounding a droplet of
Bose-Einstein condensate~--- Bose star. Our approach is based on
  the observation that kinetic equation for this gas is scale-symmetric
  if it is spatially homogeneous and gravitational scattering of
  its  particles is represented by Landau integral. Symmetry plus a
possibility of particle/energy exchanges with the condensate
provide a family of self-similar kinetic solutions~(\ref{eq:an}),
(\ref{eq:ab}) with arbitrary scaling dimensions~$D$, finite energies
and multiplicities. The latter solutions  describe gas
condensation at~${2 < D  <   3}$  and growth of its mass (evaporation
of the condensate)  at~${D>3}$. We computed profiles of
  solutions, both stable in Fig.~\ref{fig:schematic}(b) with
Rayleigh–Jeans low-energy tails ${F\propto \omega^{-1/2}}$ and
unstable with behavior~$F\propto \omega^{-1/3}$ as~$\omega\to
  0$.

Importantly, our stable scaling solutions work as ``nonthermal
attractors''~\cite{Berges:2008wm} for kinetic evolutions which
approach them regardless of the initial data. Moreover, the 
evolutions remain approximately self-similar even if scale
breaking is present  in the kinetic equation. In this new regime of
adiabatic self-similarity the distribution function stays close
to self-similar profiles as their scaling dimension~${D = D(t)}$
  drifts slowly.

Notably, we observed that adiabatic self-similarity is a powerful tool
for solving essentially non-equilibrium kinetic problems. In this
method one finds a succession of self-similar solutions with
different~$D$ and then specifies the time drift of the latter on the
basis of conservation laws. Alternatively, one may skip calculation of
scaling solutions at the cost of few heuristic constants remaining in
the approximate evolution. Acting in this way, we obtained growth
law~(\ref{eq:44}) of Bose star mass due to particle condensation from
the surrounding gas, see also Ref.~\cite{Dmitriev:2023ipv}.

Our study opens several directions for future research. First, the
approach of adiabatic self-similarity, if applied to ordinary
non-gravitational kinetics, may generalize existing ``nonthermal 
attractors''~\cite{Semikoz:1994zp, *Semikoz:1995rd,  Nowak:2012gd,
  Berges:2013eia, *Berges:2013fga, *Berges:2014bba,
  PineiroOrioli:2015cpb, Chantesana:2018qsb, Schmied:2018mte,
  Semisalov:2021} to systems with slightly broken scale symmetry. This
may explain drift of scaling parameters~\cite{Schmied:2018upn,
  Heller:2023mah} in simulations and other scale-asymmetric
phenomena, while detailed comparison of scaling solutions with
  numerical results (cf.\ Fig.~\ref{fig:com1})  will confirm
  reliability of the approach.
Second, kinetics of an inhomogeneous  gravitationally
interacting gas may honor another type of scale symmetry
involving~$\bm{x}$ transformations. If in there, it would
produce a  new class of self-similar solutions for bosonic gas
cloud (halo/minicluster) with Bose star in the center that
simultaneously shrinks in size and rescales its particle
energies. Such a regime would be fascinating to study 
or discover in simulations.

Third and finally, application of our adiabatic approach to growth
of dark matter Bose stars~\cite{Dmitriev:2023ipv} was surprisingly
successful:  crude requirement of self-similarity provided heuristic 
law~\eqref{eq:44} explaining the simulation results. This method can
be improved and derived from first principles. For a start, one can deduce
scattering integral for particle condensation onto the Bose
star, cf.~\cite{Chan:2022bkz, Dmitriev:2023jnu}, thus uncovering the last unknown scale-breaking 
term in the kinetic equation~--- analog of the source and the sponge
in Eq.~\eqref{eq:13}. With that in hand, adiabatic expansion in
Sec.~\ref{sec:apprx} would give consistent self-similar description of
condensation and correct equation for~$D(t)$. 

It is worth noting that validated comprehensive growth law for dark
matter Bose stars can disclose distribution of these objects
in the present-day Universe, cf.~\cite{Maseizik:2024qly, *Maseizik:2024uln, 
  Gorghetto:2024vnp}, thus opening new paths for the discovery of (or
constrains on) ultralight and axion-like dark matter.  


\begin{acknowledgments}
  This work is supported by the grant {RSF~22-12-00215-p}.
\end{acknowledgments}


\appendix


\section{Kinetic equation for homogeneous and isotropic gravitating gas} 
\label{sec:KE_hsse}

In spherical symmetry, homogeneous kinetic equation~\eqref{eq:LE},
\eqref{eq:si} further simplifies because distribution function~$f_{\bm 
  p}$ depends only on~${p =  |\bm{p}|}$ and Landau flux is collinear
with momentum: ${s_i  = s(p)\, p_i/p}$. We obtain
\begin{equation}
\label{eq:f_p2s}
\partial_{t} f_p  = - \partial_{p_j} s_j = - p^{-2} \partial_{p}
(p^2s)
\end{equation}
and
\[
s(p)  = \frac{G^2 m^4\Lambda}{4\pi^2p}\int \frac{d^3{\bm q}}{u} 
  \, p_i {\cal P}_{ij}\left[f_p^2 \partial_{q} f_{q}\frac{q_j}{q} -
    f_q^2\partial_{p} f_p\frac{p_j}{p}\right]\,.
  \]
Convolutions of~${{\cal P}_{ij} \equiv \delta_{ij} - u_i u_j
  /u^2}$ with $\bm{p}$ and~$\bm{q}$ depend on the angle~$\theta$
between these two vectors: 
\[
   p_ip_j{\cal P}_{ij} = p_iq_j{\cal P}_{ij} = \frac{p^2 q^2 \sin^2 \theta}{m^2 u^2}\,,
   \]
   where ~${\bm{u} \equiv (\bm{p} - \bm{q})/m}$ and~$m^2u^2 = 
p^2 - 2pq \cos\theta + q^2 $. Explicitly integrating
   over directions of~$\bm{q}$,
\[
  \int d\Omega_{\bm{q}} \; \frac{\sin^2 \theta}{m^3 u^3} =
  \frac{8\pi}{3p^3q^3} \, \min(p^3, q^3)\,, 
  \]
we recast the Landau flux in the form of one-dimensional integral
\[
s  = \frac{2G^2 m^5\Lambda}{3\pi p^3}\int\limits_0^{\infty} dq \, \min(p^3,
  q^3) \, \left(pf_p^2 \partial_q f_q - q f_q^2 \partial_p f_p
  \right)\,,
  \]  
where~${d\Omega_{\bm{q}} \equiv 2\pi  d(\cos \theta)}$. Next, we open 
  the brackets and integrate the first term by parts,
\begin{equation}
  \label{eq:KE_f}
  s  =  - \frac{2G^2m^5 \Lambda}{\pi p^2} \left( B_p f_p^2 + A_p
  \partial_{p} f_p\right)  = p^{-3} \partial_p W_p\,,
\end{equation}
where 
\begin{equation}\label{eq:AB_f}
A_p \equiv \int\limits_0^{\infty} \frac{q dq}{3p} \;
\mathrm{min}(p^3,q^3) f_q^2 \,,\quad 
B_p \equiv \int\limits_0^{p} q^2 dq\,  f_q\,,
\end{equation}
and we noticed that the right-hand side of Eq.~\eqref{eq:KE_f} is
proportional to the total derivative of the preflux
\begin{equation}
  \label{eq:47}
  W_p = \frac{2G^2 m^5 \Lambda}{\pi} (B_p C_p - p A_p f_p)\,, \;\;
  C_p = \int\limits_p^{\infty} q dq \, f_q^2\,,
\end{equation}
see Eqs.~(\ref{eq:AB_f}) implying, in particular, $\partial_p A_p
  \equiv pC_p -A_p / p$.

The last step is to change notations in
Eqs.~(\ref{eq:f_p2s}), (\ref{eq:KE_f}), (\ref{eq:AB_f}), and
(\ref{eq:47}): introduce distribution~$F(t,\, \omega)$ of
  particles over
energies~${\omega \equiv p^2 / 2m}$ in Eq.~\eqref{eq:9}, total
particle flux
\[
  {{\cal J}_N \equiv 4\pi p^2 s\, V_R/(2\pi)^3}\,,
  \] rescaled preflux~${W
\equiv W_p V_R/2m\pi^2 }$, and integrals $A \equiv A_p\,
  mV_R^2/4\pi^4$, $B \equiv B_p \,V_R/2\pi^2$, ${C \equiv C_p \, m^2
V_R^2/(4\pi^4)}$. We arrive at kinetic equation~\eqref{eq:14} from the
main text; note that the energy flux~${\cal J}_E$ is defined in
Eqs.~\eqref{eq:2}, \eqref{eq:8}.


\section{Numerical methods for nonstationary kinetics}
\label{sec:numer-solut-time}

Let us solve the time-dependent kinetic equation~\eqref{eq:13} numerically.
In Apps.~\ref{sec:numer-solut-time} and~\ref{sec:self-simil-prof}
we work in dimensionless  units~${N_0 = 2\omega_0 = t_{\mathrm{rel}} =
  1}$, Eq.~\eqref{eq:12}, but omit tildes above the quantities.

We introduce uniform grid ${\omega_{j} = j\Delta \omega}$ in the
energy domain, where ${0 \leq j \leq N-1}$ indexes sites and 
$\Delta \omega = \omega_{\mathrm{max}}/(N-1)$ is
lattice spacing. The cutoff ${\omega_{\mathrm{max}} = 30}$ is
chosen large enough to make ${F(\omega_{\mathrm{max}})\lesssim
  10^{-50}}$ negligibly small throughout the simulations.
Accordingly, we impose Dirichlet conditions
\begin{equation}
  \label{eq:48}
  F_0 = F_{N-1} = 0
\end{equation}
at the lattice boundaries. The size of our lattice size belongs
to the
  interval~${2\cdot 10^4  \leq N \leq 6\cdot 10^4}$ which
  gives~${\Delta \omega \sim 10^{-3}}$ and  percent-level
discretization errors in the numerical scheme below. 

We discretize the collision integral~\eqref{eq:14}
  using central difference
\begin{equation}
\label{eq:Stf}
\mathrm{St}\, F_j = -(W_{j+1} + W_{j-1} -
  2 W_j )/\Delta \omega^2\,,
\end{equation}
and trapezoidal rule for the basic integrals~$A$,~$B$,
and~$C$, e.g.,~${B_j = \Delta \omega \sum_{j'=1}^{j-1}
  F_j' + \Delta \omega F_j/2}$. Time steps~$\Delta t^n$ are
  provided by leap-frog method,
\begin{equation}
  \label{eq:scheme}
  \frac{F_j^{n+1} - F_j^{n}}{\Delta t^n} = \mathrm{St}\,
  F_j^{n+1/2}
  - \mu_j F^{n+1/2}_j + J_{\mathrm{ext}, j}^{n+1/2}\,,
\end{equation}
where~$n$ indexes the time sites,~${\Delta t^n = 
  t^{n+1} - t^n}$, we introduced short-hand notations
${F^{n}_j =  F(t^n,\, \omega_j)}$, ${\mu_j = \mu(\omega_j)}$,
${J_{\mathrm{ext}, j}^{n+1/2} = J_{\mathrm{ext}}(t^{n+1/2},\,
  \omega_j)}$, and the right-hand side is computed at the
midpoint values ${F_j^{n+1/2} \equiv (F_j^{n} + F_j^{n+1})/2}$ and
${t^{n+1/2} \equiv (t^{n}+ t^{n+1})/2}$. This gives second-order
discretization scheme  with errors of order~${O(\Delta t^2, \Delta
  \omega^2)}$.

We use the sponge~\eqref{eq:17} with
thickness~${\omega_{\mathrm{IR}} \sim \Delta \omega}$ comparable to
one grid spacing~--- to minimize its impact at
finite~$\omega$. On the other hand, the amplitude of our sponge is
relatively large, ${\mu_0 =  2\cdot 10^6}$, so it efficiently absorbs
particles with~${\omega\approx 0}$.

At every time step, we solve finite-difference
equation~\eqref{eq:scheme} with boundary conditions~(\ref{eq:48})~---
a nonlinear algebraic system for~${N-2}$ unknowns~$F_j^{n+1}$. To
this end we exploit iterated Crank-Nicholson
method~\cite{Teukolsky:1999rm}. At the zeroth iteration, we substitute
the guess~${F^{n+1}_i = F^{n}_i}$ into the right-hand side of
Eq.~(\ref{eq:scheme}) thus turning it into an explicit 
  formula for the first-iteration
unknowns~$(F_j^{n+1})_1$. Finding the latter and substituting
them into the right-hand side, again, we get the 
second-iteration unknowns~$(F_j^{n+1})_2$, etc. We  checked that small
step~${\Delta t   \lesssim \Delta \omega^2}$ and precisely three 
iterations are sufficient to ensure second-order precision and
numerical stability, cf.\ Ref.~\cite{Teukolsky:1999rm}.   

To handle fast-evolving distributions, we implement adaptive
time stepping. Prior to every iteration, we estimate by
linear extrapolation the step $\Delta
t_{\mathrm{dyn}}^n$ needed for~1\% relative change
between~$F_j^{n}$ and~$F_j^{n+1}$. The actual step is then chosen
as~${\Delta t^n = \mathrm{min}(\Delta t_{\mathrm{dyn}}^n,\,
  \Delta \omega^2/4)}$ 
to ensure both accuracy and stability. This adaptive technique
allows us to employ strong sponges with large~$\mu_0$: the solution
cannot be altered at once by more than~1\% anyway.  

We estimate discretization errors by varying~$\Delta t$ and~$\Delta
\omega$. This reveals that our solutions are stable
on~$0.5\%$ precision level in the entire domain of~${\tau
  \lesssim 150}$ relaxation times. Narrow and strong
sponge~\eqref{eq:17} ensures that the energies of solutions drift at
most by~$1 \%$ while multiplicities change essentially:
${N(150) \sim (0.1 \, \div \,1) \, N(0)}$. This justifies usage of
energy conservation in Sec.~\ref{sec:compare}.


\section{Numerics for self-similar profiles}
\label{sec:self-simil-prof}

We solve integro-differential profile equation~(\ref{eq:4}),
(\ref{eq:14}) by recasting it as a system of  ordinary
differential equations. The latter are supplemented with 
boundary conditions: one of Eqs.~\eqref{eq:1},~\eqref{eq:25},
or~\eqref{eq:33} at low~$\omega_s$ and a requirement of fast
falloff at infinity,
  \begin{equation}
    \label{eq:46}
    F_s(\omega_{\mathrm{UV},s}) = C_s(\omega_{\mathrm{UV},s}) = 0 \,, \qquad 
    \omega_{\mathrm{UV},s} \to +\infty\,,
  \end{equation}
  where the second equality follows from the definition of~$C_s$ in
  Eqs.~\eqref{eq:ABC}. In this Appendix we adopt dimensionless
  units~\eqref{eq:12} but do not write tildes, again. 
  
It is straightforward to see that indefinite
integrals $A_s(\omega_s)$, $B_s(\omega_s)$, and $C_s(\omega_s)$ in
Eqs.~\eqref{eq:ABC} satisfy equations
  \begin{align}
  \label{eq:NumABC}
  & \partial_{\omega_s}A_s = - A_s/(2\omega_s) + C_s \,, 
  \quad \partial_{\omega_s}B_s = F_s\,,\\ \notag
  &\partial_{\omega_s}C_s = -F_s^2/(2\omega_s)\,.
\end{align}
  Next, substituting~$\mathrm{St}\, F_s$ from Eq.~\eqref{eq:2}
  into Eq.~(\ref{eq:4}) and reshuffling the derivatives, we find,
  \begin{equation}
    \label{eq:49}
    \partial_{\omega_s} {\cal J}'_{N,s} = -k_NF_s+J_{\mathrm{ext},s}\,,
  \end{equation}
  where~${k_N = 1-3/D}$ and 
  \begin{equation}
    \label{eq:51}
    {{\cal J}'_{N,s}(\omega_s) \equiv {\cal J}_{N,s} +
      k_\beta\omega_sF_s}\,, \quad k_\beta\, \equiv 2/D-1\,.
  \end{equation}
  Finally, we explicitly evaluate the derivatives of the 
  preflux~\eqref{eq:10} in Eq.~\eqref{eq:8} and get
  \begin{multline}
    \label{eq:Num2}
   \partial_{\omega_s} F_s = \frac{1}{A_s}
   \bigg[\frac{F_s}{2\omega_s} (A_s - B_s F_s) -
     {\cal J}'_{N,s}  + k_\beta\omega_sF_s \bigg],
  \end{multline}
  where the right-hand side is expressed in terms of~${\cal J}_{N,s}'$. 

Relations~(\ref{eq:NumABC}), \eqref{eq:49}, and~\eqref{eq:Num2} form 
a system of five first-order differential equations for the
unknowns~$F_s(\omega_s)$, ${\cal J}_{N,s}'(\omega_s)$,
$A_{s}(\omega_s)$, $B_{s}(\omega_s)$, and~$C(\omega_s)$. The simplest
option is to impose the boundary condition~\eqref{eq:25} at low
$\omega_s$: ${F_s(\omega_s) = 0}$ at~${\omega_s \leq
  \omega_{\mathrm{IR},s}}$, see also Eq.~\eqref{eq:40}. This gives
\begin{equation}
  \label{eq:BC_0}
  F_s = B_s = 0 \;\; \mbox{and}\;\;
  A_s = \frac{2}{3}\omega_{\mathrm{IR},s} C_s \;\; \mbox{at} \;\;
  \omega_s = \omega_{\mathrm{IR},s} .
\end{equation}
We thus deduced five boundary conditions~(\ref{eq:46}),
\eqref{eq:BC_0} for the same number of equations. 

We solve the latter by shooting method. For a start, using~${\cal
  J}_{N,s}'(\omega_{\mathrm{IR},s})$,~$C_s(\omega_{\mathrm{IR},s})$, and
Eqs.~\eqref{eq:BC_0} as initial data, we numerically evolve
Eqs.~(\ref{eq:NumABC}), \eqref{eq:49}, and~\eqref{eq:Num2} from~${\omega_s
  = \omega_{\mathrm{IR},s}}$ to~${\omega_s = \omega_{\mathrm{UV},s}}$. After that
we tune the values of~${\cal J}_{N,s}'(\omega_{\mathrm{IR},s})$
and~$C_s({\omega_{\mathrm{IR},s}})$ to make
$|C_s(\omega_{\mathrm{UV},s})|$ and~${|F_s(\omega_{\mathrm{UV},s})|}$ smaller
than~$10^{-12}$ in accordance with the falloff
conditions~(\ref{eq:46}). This provides the profile~$F_s(\omega_s)$, 
the integrals~$A_s$, $B_s$, $C_s$, and particle flux~${\cal
  J}_{N,s}(\omega_s)$ in Eq.~\eqref{eq:51} for given~$D$ and~$J_{\mathrm{ext},
  s}(\omega_s)$. 

The last step is to compute conserved charges. Rescaled particle
number~$N_s$ is given by the conservation law~\eqref{eq:N_J}, where
condensation flux equals~${{\cal J}_{N,s}(0) = {\cal
    J}_{N,s}'(\omega_{\mathrm{IR},s})}$ and the integral\footnote{We
compute it by adding~${\partial_{\omega_s} I(\omega_s) =
  J_{\mathrm{ext},s}}$ to the system of differential equations, 
where~${I(\omega_{\mathrm{IR},s}) =0}$.} of~$J_{\mathrm{ext},s}$ now
runs from~$\omega_{\mathrm{IR},s}$
to~$\omega_{\mathrm{UV},s}$. At finite~$\omega_{\mathrm{IR},s}$, the
energy conservation law includes the boundary term, 
\begin{equation}
  \label{eq:50}  
  k_EE_{s}= {\cal J}_{E,s} (\omega_{\mathrm{IR},s}) + 
  \int\limits_{\omega_{\mathrm{IR},s}}^{\omega_{\mathrm{UV},s}} \omega_s
  J_{\mathrm{ext},s}\, d\omega_s \,,
\end{equation}
cf.\ Eq.~\eqref{eq:E_J}. Extracting~${\cal J}_{E,s}
(\omega_{\mathrm{IR},s}) = \omega_{\mathrm{IR},s} {\cal
  J}_{N,s}'(\omega_{\mathrm{IR},s})$ from Eqs.~\eqref{eq:8} and
numerical solution, we get~$E_s$.

In practice, we set~$\omega_{\mathrm{UV},s} = 20$ and
vary~$\omega_{\mathrm{IR},s}$ in the interval~${10^{-8} \leq
    \omega_{\mathrm{IR},s} \leq 10^{-2}}$. We evolve equations
using fourth-order Runge-Kutta method with step size~${\Delta
  \omega_s = \omega_{\mathrm{IR},s}/10}$ or adaptive
Bulirsch-Stoer algorithm~\cite{NR} with target precision~${\Delta F_s =
    10^{-11}}$. Numerical errors are controlled by changing lattice 
  parameters and comparing conservation laws~\eqref{eq:N_J}
  and~\eqref{eq:50} with charge definitions~\eqref{eq:ENs}. Thus
  estimated, precision of our profiles is always better  
  than~${\Delta \ln F_s < 5\cdot 10^{-7}}$. Finally, we ensure 
  that all solutions have correct ultraviolet   behavior~\eqref{eq:30} and
  satisfy Eq.~\eqref{eq:1} in the limit $\omega_{\mathrm{IR},s} \to 0$.  

Another option is to impose asymptotic (${\omega_s= 0}$)
conditions~\eqref{eq:1} or~\eqref{eq:33} from the start. To this end we
reformulate them as requirements at small but finite $\omega_s =
  \omega_{\mathrm{IR},s} \sim 10^{-8}$. 
Namely, at~${\omega_s \leq
  \omega_{\mathrm{IR},s}}$ we fix one of the two asymptotic 
forms~\eqref{eq:34} or~\eqref{eq:31} of solutions, 
\begin{equation}
  \label{eq:54}
  F_s = F_0\,\omega_{s}^{-1/2} + F_1 -
  \frac{F_1^2}{2F_0}\, \omega_{s}^{1/2}\;\;\,  \mbox{or} \;\;\,
  F_s = F_0\,\omega_{s}^{-1/3},  
\end{equation}
where the constants~$F_0$ and~$F_1$ in the first case control particle
and energy fluxes at~$\omega_{s}=0$, in particular, ${\cal J}_{E,s}(0)
= -3 F_0^2 F_1/2$. In the second case~${{\cal J}_{E,s}(0)}$ is not
free, see discussion in App.~\ref{sec:gamma=-1/3}. 
Next, we determine four Cauchy data~$F_s(\omega_{\mathrm{IR},s})$,
$A_s(\omega_{\mathrm{IR},s})$, $B_s(\omega_{\mathrm{IR},s})$,
and~$C_s(\omega_{\mathrm{IR},s})$ from Eqs.~\eqref{eq:54}
and~\eqref{eq:ABC}. In particular,
\begin{equation}
  \notag
  C_s(\omega_{\mathrm{IR},s}) = \frac{F_0^2}{2\,\omega_{\mathrm{IR},s}} +
  \frac{2F_0F_1}{\omega_{\mathrm{IR},s}^{1/2}} + C_0 \;\;\; \mbox{or}\;\;\;
  \frac{3F_0^2}{4\,\omega_{\mathrm{IR},s}^{2/3}} +  C_0,
\end{equation}
and
\begin{align}
  \notag
  &A(\omega_{\mathrm{IR},s})  = \frac{2}{3}\big[F_0^2 + F_0F_1\,
  \omega_{\mathrm{IR},s}^{1/2} + \omega_{\mathrm{IR},s}
  C_s(\omega_{\mathrm{IR},s})\big]\\
  \notag 
  &\qquad \mbox{or} \;\;\;\; \frac{2}{5}F_0^2 \, \omega_{\mathrm{IR},s}^{1/3}
  +\frac{2}{3}\omega_{\mathrm{IR},s}C_s(\omega_{\mathrm{IR},s})\,.
\end{align}
We are left with three undetermined\footnote{One can
    parametrize solutions by rescaled particle number~$N_s$ or
    source amplitude~$J_0$ instead of ${\cal
      J}_{N,s}'(\omega_{\mathrm{IR},s})$ finding the latter from the
    conservation law~\eqref{eq:N_J} and Eq.~\eqref{eq:51}.}
parameters~$F_0$,~$C_0$, and ${\cal
  J}_{N,s}'(\omega_{\mathrm{IR},s})$: recall that the extra
coefficient~$F_1$ in Eq.~\eqref{eq:54} is related to~${\cal 
  J}_{E,s}(0)$ and fixed by boundary conditions~\eqref{eq:1}
or~\eqref{eq:33}. 

But this is not the end of the story because parameter~$C_0$ is
also not free. In Appendix~\ref{sec:asympt-as-omeg} we relate it 
to~${\cal J}_{N,s}'(\omega_{\mathrm{IR},s})$, via
Eqs.~\eqref{eq:51},~\eqref{eq:59} or Eqs.~\eqref{eq:51},~\eqref{eq:60}
in the two cases given above. Once~$C_0$ is fixed, the actual
shooting parameters are~$F_0$ and~${\cal
  J}_{N,s}'(\omega_{\mathrm{IR},s})$. Tuning them to solve the 
falloff conditions~\eqref{eq:46}, we obtain one numerical 
solution for every~${\cal J}_{E,s} (0)$, $D$, and ${\cal
  J}_{\mathrm{ext},s}(\omega_s)$.

Checking sensitivity of results to~$\omega_{\mathrm{IR},s}$
we measure numerical accuracy of imposing asymptotic
conditions~\eqref{eq:1} or~\eqref{eq:33}  which never exceeds ${\Delta
  \ln F_s < 10^{-6}}$.

We finish this Section with comments on numerical solutions
at~${J_{\mathrm{ext}, s}   = 0}$ and~${D=5/2}$. The respective
profile equation~\eqref{eq:4} has residual scale
symmetry~\eqref{eq:24} if zero-flux condition~\eqref{eq:1} is
  imposed. We numerically observed   
that adjustment of one shooting parameter~$F_0$ in this case is 
sufficient to satisfy both falloff conditions~\eqref{eq:46} at
once. The second free parameter then normalizes solutions,
say,~${{\cal J}_{N,s}(0) = - 1/2}$.

A somewhat different case is~${J_{\mathrm{ext},s}=0}$ combined with
condition~\eqref{eq:25} at finite~$\omega_{\mathrm{IR},s}$. The 
latter slightly breaks the residual symmetry~\eqref{eq:24} ascertaining that
both~$F_0$ and~${{\cal J}_{N,s}'(\omega_{\mathrm{IR},s})}$
should be tuned to
  satisfy the falloff conditions. Besides,
  finite~$\omega_{\mathrm{IR},s}$ leads to tiny departure of~$D$ 
  from~$5/2$. Indeed,  
  energy conservation~\eqref{eq:50} reads,
  \begin{equation}
    \label{eq:63}
    (2-5/D) \, E_s = {\cal J}_{E,s}(\omega_{\mathrm{IR},s})  =
    \omega_{\mathrm{IR},s} {\cal J}_{N,s}'(\omega_{\mathrm{IR},s}) \,,
  \end{equation}
  where the last equality follows from Eq.~\eqref{eq:8} and boundary
  conditions~\eqref{eq:BC_0}. One can therefore use~${\cal
    J}_{N,s}'(\omega_{\mathrm{IR},s})$ as a shooting parameter
  and find~$D$ from Eq.~(\ref{eq:63}), or, better, fix the  
  particle flux and satisfy the falloff conditions adjusting 
  a tiny difference~${D-5/2}$. 


\section{Asymptotics of self-similar profiles}
\label{sec:asympt-self-simil}


\subsection{Asymptotics as \texorpdfstring{$\omega_s \rightarrow 0$}{}}
\label{sec:asympt-as-omeg}

In Sec.~\ref{sec:absence-power-law} we identified
two\footnote{Behaviors~$F_s \propto \omega_s^{-3/4}$ and~$\omega_s^0$
are subtle: they produce logarithmic divergences and~${\ln
\omega_s}$ terms in the collision integral,
cf.\ Eqs.~\eqref{eq:ABC}. These inconsistencies do not
  disappear even after multiplying the asymptotics 
with~$(\ln\omega_s)^{\zeta}$. We thus exclude subtle
behaviors from consideration.} 
nontrivial low-energy behaviors~${F_s \propto \omega_s^{-1/2}}$
and~$\omega_s^{-1/3}$ that  make collision integral apparently infrared
convergent and support finite fluxes as~$\omega_s\to 0$.  Below we
build power-law expansions of the profiles on the basis of these
behaviors. We  will consider them separately.


\subsubsection{Behavior~\texorpdfstring{${F_s \propto \omega_s^{-1/2}}$}{}
  at low~\texorpdfstring{$\omega_s$}{}}
\label{sec:gamma=-1/2}

It is natural to assume half-integer power-law expansion for such
profiles,
\begin{equation}
  \label{eq:52}
  F_s = F_0\, \omega_s^{-1/2}+ F_1 + F_2\, \omega_s^{1/2} + F_3\,
  \omega_s + O(\omega_s^{3/2})\,,
\end{equation}
where~$F_i$ are constants. But this representation immediately
gives~$\ln \omega_s$ in the integral
\begin{multline}
  \label{eq:55}
  C_s(\omega_s)= C_0 +
  \frac{F_0^2}{2\omega_s}+\frac{2F_0F_1}{\omega_s^{1/2}} \\-
  \left(\frac{F_1^2}{2}+F_0F_2\right)\ln\omega_s+O(\omega_s^{1/2})\,, 
\end{multline}
where~$C_0$ is the integration constant, see Eq.~\eqref{eq:ABC}. 
Such logarithmic terms proliferate in the  full scattering
integral preventing the power-law Ansatz~(\ref{eq:52}) 
from passing the profile equation and~--- importantly~--- making
the particle flux~\eqref{eq:8}, \eqref{eq:10} diverge in the infrared:
${{\cal J}_{N,s} = - \frac13W_0 F_0\, (F_1^2  +2F_0 F_2)
  \, \omega_s^{-1/2} \ln \omega_s} + \dots$. This leaves only one way to
satisfy the finite-flux condition: cancel logarithms from the
get-go by setting
\begin{equation}
  \label{eq:53}
  F_2 = -\frac{F_1^2}{2F_0}\,.
\end{equation}
Relation~\eqref{eq:53} is taken into account in Eqs.~\eqref{eq:32}
and~\eqref{eq:34} of the main text.

Substituting expansion~(\ref{eq:52}), \eqref{eq:53} into 
other integrals\footnote{Note that~$A_s 
\equiv \frac{2}{3} \omega_s C_s(\omega_s) + \frac13\int_{0}^{\omega_s}
d\omega_s' \, (\omega_s'/\omega_s)^{1/2} F_s^2(\omega_s')$ 
includes the same integration constant~$C_0$.}~\eqref{eq:10},
\eqref{eq:ABC}, we arrive at the preflux  
\begin{multline}
  \label{eq:56}
  \frac{W_s}{W_0} = \frac{3}{2} F_0^2 F_1
  + \frac{\omega_s^{1/2}}{3} (4 F_0C_0 + F_0F_1^2)  \\
  + \frac{\omega_s}{12} (4 F_1 C_0 + 22 F_1^3 - 45 F_0^2 F_3) + O(\omega_s^{3/2})
\end{multline}
and hence at the scattering integral~\eqref{eq:2}, 
\begin{equation}
\label{eq:-1/2b}
\mathrm{St} F_s= \frac{W_0}{12}(4 F_0 C_0 + F_0
F_1^2)\,\omega_s^{-3/2}+O(\omega_s^{-1/2})\,.
\end{equation}
Notably, the leading-order term of~$\mathrm{St}\, F_s$ had already
disappeared, cf.\ Eq.~\eqref{eq:StFcascade}. But the
remaining terms still dominate over the~$O(\omega_s^{-1/2})$ 
left-hand side of  the profile equation~\eqref{eq:4}. To satisfy it, we impose
condition
\begin{equation}
  \label{eq:57}  
  C_0 \equiv  \left[C_s(\omega_{\mathrm{IR},s})
  -\frac{F_0^2}{2\omega_{\mathrm{IR},s}}
  -\frac{2F_0F_1}{\omega_{\mathrm{IR},s}^{1/2}}
  \right]_{\omega_{\mathrm{IR},s} \to 0}
  \!\!\!\!\!\!\!\!\!\!= -\frac{F_1^2}{4},
\end{equation}
where the first equality is the definition of~$C_0$ following from
Eq.~\eqref{eq:55} and the second is
needed to cancel the~$\omega_s^{-3/2}$ term in Eq.~(\ref{eq:-1/2b}).

Evaluating the derivatives of the preflux~(\ref{eq:56}), we get
finite fluxes at~${\omega_s=0}$,
\begin{align}
\notag
&{\cal J}_{N,s}(0) \equiv \partial_{\omega_s} W_s(0) =\frac{7}{4} \,
W_0F_1^3-\frac{15}{4}\, W_0F_0^2F_3\,, \\ 
&{\cal J}_{E,s}(0) \equiv -W_s(0) =-\frac{3}{2}\, W_0F_0^2F_1\,, \notag
\end{align}
where Eq.~(\ref{eq:57}) was used. These expressions are helpful
Sec.~\ref{sec:self-simil-solut}, cf.\ Eqs.~\eqref{eq:32} and~\eqref{eq:34}.

Once Eqs.~\eqref{eq:53} and~\eqref{eq:57} are met, the
left- and right-hand sides of the profile equation~\eqref{eq:4} become
comparable. Solving it  order-by-order, one determines~$F_4$, $F_5$,
$F_6$, etc. Constants~$F_0$, $F_1$, and~$F_3$ of the asymptotics
  are free, where~$F_1$ controls energy flux at~$\omega_s=0$
and~$F_0$,~$F_3$ should be tuned to satisfy the falloff
conditions~(\ref{eq:46}). We thus have one solution for every~$D$,
$J_{\mathrm{ext},s}(\omega_s)$, and~${\cal J}_{E,s}(0)$.

We finish this subsection with a remark. In the numerical method of
App.~\ref{sec:self-simil-prof} we determined~$C_0$ from the leading
asymptotics of the particle flux,
\begin{equation}
  \label{eq:59}
        {\cal J}_{N,s}(\omega_s) = \frac{\omega_s^{-1/2}}{6} \, W_0 F_0 ( 4
        C_0 + F_1^2) + O(\omega_s^{0})\,,
\end{equation}
where Eqs.~\eqref{eq:8},~(\ref{eq:53}), \eqref{eq:56} were used
  and Eq.~\eqref{eq:57} was not. For finite~${{\cal
      J}_{N,s}}$ as~${\omega_s\to 0}$, this definition of~$C_0$ is equivalent to
Eq.~\eqref{eq:57}.


\subsubsection{Behavior~\texorpdfstring{${F_s\propto \omega_s^{-1/3}}$}{}
  at low~\texorpdfstring{${\omega_s}$}{}}
\label{sec:gamma=-1/3}

This time, we start with power counting. Suppose low-energy
asymptotics of the profile looks like
\[
F_s(\omega_s)  = 
F_0\, \omega_s^{-1/3}+F_1\, \omega_s^{\zeta}+o(\omega_s^{\zeta})\,,
\quad\zeta > -1/3\,.
\]
Then the preflux~\eqref{eq:10}, \eqref{eq:ABC} includes powers
\begin{align}
\label{eq:-1/3b}
W_s= d_0 + d_{1/3}\,  \omega_s^{2/3} +  d_{\zeta} \,
\omega_s^{\zeta+1/3} + o(\omega_s^{\zeta+1/3})\,,
\end{align}
with coefficients~$d_0$,~$d_{1/3}$, and~$d_{\zeta}$. The respective
scattering integral~${\mathrm{St}\, F_s = - \partial_{\omega_s}^2 W_s
  = \frac{2}{9}\, d_{1/3}\, \omega_s^{-4/3}} + \dots$ receives no contribution from
the leading-order~$d_0$ term, but still cannot be compensated by 
the~$O(\omega_s^{-1/3})$ left-hand side of the profile
equation~\eqref{eq:4}. Hence, we require~${d_{1/3} = 0}$ and restrict
subdominant terms to have~${\zeta = 2/3}$ and~${\zeta \geq
4/3}$. Indeed,~$\zeta=2/3$ gives~$W_s = d_\zeta \,\omega_s$ which
  does not affect the scattering integral~\eqref{eq:2},
\eqref{eq:8}. At $\zeta \geq 4/3$ the integral~$\mathrm{St}\, F_s \lesssim
\omega_s^{-1/3}$ can be balanced in the profile equation.

We ended up with asymptotics of the form
\begin{equation}
  \label{eq:61}
  F_s(\omega_s) = F_0\, \omega_s^{-1/3}+
  F_1\, \omega_s^{2/3}+O(\omega_s^{4/3})\,,
\end{equation}
cf.\ Eq.~\eqref{eq:31} from the main text. Unlike in the previous
subsection, the integrals~\eqref{eq:ABC} do not include logarithms, e.g.,
\begin{equation}
  \label{eq:58}
  C_s= \frac{3}{4} \,  F_0^2 \omega_s^{-2/3} + C_0 -3 F_0 F_1\, \omega_s^{1/3}
    + O(\omega_s^{4/3})
\end{equation}
with integration constant~$C_0$. We get the preflux~\eqref{eq:10},
 \[
 \frac{W_s}{W_0} = \frac{9}{40} F_0^3 + \frac{5}{6} F_0 C_0
 \,\omega_s^{2/3} - \frac{729}{220} F_0^2 F_1\,  \omega_s + O(\omega_s^{5/3})\,,
 \]
where the second term generates the leading power-law
 in~${\mathrm{St}\, F_s = - \partial_{\omega_s}^2 W_s \propto  C_0
 F_0 \, \omega_s^{-4/3}}$. As discussed, we get rid of it by setting
 \begin{equation}
   C_0 \equiv \left[ C_s(\omega_{\mathrm{IR},s})  -
     \frac{3F_0^2}{4\omega_{\mathrm{IR},s}^{2/3}} 
   \right]_{\omega_{\mathrm{IR},s}\to 0} = 0\,,
 \end{equation}
 where the first equality is a definition of~$C_0$ in Eq.~(\ref{eq:58}).
 
We again evaluate condensation fluxes~--- derivatives~\eqref{eq:8}
of~$W_s(\omega_s)$ at~$C_0 = \omega_s=0$,
\[
{\cal J}_{N,s}(0)=-\frac{729}{220}W_0F_0^2F_1\,, \quad 
{\cal J}_{E,s}(0)=-\frac{9}{40}W_0 F_0^3\,.
\]
These expressions are discussed in Sec.~\ref{sec:self_eq} below
Eq.~\eqref{eq:31}. Let us also write down the leading-order
particle flux at~${C_0 \ne  
  0}$,
\begin{equation}
  \label{eq:60}
  {\cal J}_{N,s}(\omega_s) = \frac{5}{9}W_0 F_0 C_0\, \omega_s^{-1/3} + O(\omega_s^{0})\,,
\end{equation}
which turns useful in App.~\ref{sec:self-simil-prof}.

Setting~$C_0 = 0$ and solving the profile equation order-by-order
in~$\omega_s$, one finds all expansion coefficients in 
Eq.~\eqref{eq:61} except  for~$F_0$ and~$F_1$ which are fixed by the 
falloff conditions~(\ref{eq:46}). This gives one solution for every~$D$
and~$J_{\mathrm{ext},s}(\omega_s)$. 

Notably, our~$\omega^{-1/3}$ solutions are devoid of one free parameter 
as compared to~$\omega^{-1/2}$ profiles and therefore can be 
considered as fine tuned.


\subsection{Asymptotics as \texorpdfstring{$\omega_s \to \infty$}{}}
\label{sec:exponential_infty}

Now, we deduce high-energy behavior of localized self-similar
profiles. We will assume that they decrease exponentially 
with~$\omega_s$ thus ignoring~$F_s^2(\omega_s)$ and~$C_s(\omega_s)$
in comparison  to~$F_s$ itself. On the other hand,~${B_s \to N_s}$
and~${A_s \to G_0 /(3\,\sqrt{\omega_s})}$ as~${\omega_s \to \infty}$,
where~$N_s$ and
\[
G_0\equiv \int_0^{\infty} d\omega_s'
F_s^2(\omega_s')\sqrt{\omega_s'}
\]
are constants, cf.\ Eqs.~\eqref{eq:ABC}.

Integrating the left- and right-hand sides of the profile
equation~\eqref{eq:4},~\eqref{eq:2} from~$\omega_s$ to infinity, we
find
\begin{equation}
\label{eq:3inftyas}
k_NN_s-k_NB_s(\omega_s)= t_{\mathrm{rel}}\, {\cal J}'_{N,s}(\omega_s)
\end{equation}
and
\begin{align}
  \notag
      {\cal J}_{N,s}'(\omega_s) & \equiv {\cal J}_{N,s}(\omega_s)
        +\frac{k_\beta}{t_{\mathrm{rel}}}\,\omega_sF_s\\
  \label{eq:J_Nexp}
        & = \frac{ W_0
   G_0}{6\omega_s^{3/2}} \left( F_s - 
 2 \omega_s \partial_{\omega_s} F_s
 \right)+\frac{k_\beta}{t_{\mathrm{rel}}}\, \omega_sF_s \,,  
\end{align}
where~${k_\beta = 2/D-1}$, ${k_N = 1-3/D}$ and we
applied Eqs.~\eqref{eq:14} in the last equality.

Since~$F_s = \partial_{\omega_s} B_s$, relations~(\ref{eq:3inftyas}),
(\ref{eq:J_Nexp}) give second-order linear differential equation
for~$B_s(\omega_s)$. We solve it at high~$\omega_s$ using the WKB
Ansatz
\begin{equation}
  \label{eq:62}
  B_s(\omega_s) = N_s - B_1(\omega_s)\,\mathrm{e}^{-S_s (\omega_s)}\,,
\end{equation}
where the exponent~$S_s \gg 1$ and prefactor~$B_1$ are powers
of~$\omega_s$. To the leading~$\omega_s$ order, Eqs.~(\ref{eq:3inftyas}),
(\ref{eq:J_Nexp}) reduce to
\[
\frac{t_{\mathrm{rel}} W_0 G_0}{3\sqrt{\omega_s}}\,
\partial_{\omega_s}S_s(\omega_s)+k_\beta\omega_s =0
\]
and give solution ${S_s= - 6 k_\beta\omega_s^{5/2} /(5t_{\mathrm{rel}} W_0
G_0)}$, where we ignored the additive integration constant. In the next
order, we find equation for~$B_1(\omega_s)$,
\[
(2D-5)B_1(\omega_s)= - \omega_s(D-2)\partial_{\omega_s}B_1(\omega_s)\,,
\]
with solution $B_1= B_{0}\,\omega_s^{(5-2D)/(D-2)}$.

Finally, we substitute~$S_s(\omega_s)$, $B_1(\omega_s)$ into
Eq.~\eqref{eq:62}, draw~$W_0$ from Eq.~\eqref{eq:10}, and take
the  derivative ${F_s = \partial_{\omega_s} B_s}$. We arrive at
high-$\omega_s$ asymptotics~\eqref{eq:30} of the profile,
where $k_\beta$ and $k_N$ are expressed via $D$ and we
redefined the normalization constant as ${F_\infty \equiv - 3 k_\beta
  B_0 / (t_{\mathrm{rel}} W_0 G_0)}$. 


\bibliographystyle{apsrev4-1}
\bibliography{refs} 

@article{Jain:2023tsr,
    author = "Jain, Mudit and Wanichwecharungruang, Wisha and Thomas, Jonathan",
    title = "{Kinetic relaxation and nucleation of Bose stars in self-interacting wave dark matter}",
    eprint = "2310.00058",
    archivePrefix = "arXiv",
    doi = "10.1103/PhysRevD.109.016002",
    journal = "Phys. Rev. D",
    volume = "109",
    number = "1",
    pages = "016002",
    year = "2024"
}

@article{Schive:2014dra,
    author = "Schive, Hsi-Yu and Chiueh, Tzihong and Broadhurst, Tom",
    title = "{Cosmic Structure as the Quantum Interference of a Coherent Dark Wave}",
    eprint = "1406.6586",
    archivePrefix = "arXiv",
     doi = "10.1038/nphys2996",
    journal = "Nature Phys.",
    volume = "10",
    pages = "496--499",
    year = "2014"
}

@article{Schive:2014hza,
    author = "Schive, Hsi-Yu and Liao, Ming-Hsuan and Woo, Tak-Pong and others",
    title = "{Understanding the Core-Halo Relation of Quantum Wave
    Dark Matter from 3D Simulations}",
    eprint = "1407.7762",
    archivePrefix = "arXiv",
    doi = "10.1103/PhysRevLett.113.261302",
    journal = "Phys. Rev. Lett.",
    volume = "113",
    number = "26",
    pages = "261302",
    year = "2014"
}

@article{Liao:2024zkj,
    author = "Liao, Pin-Yu and Su, Guan-Ming and Schive, Hsi-Yu and others",
    title = "{Deciphering the Soliton-Halo Relation in Fuzzy Dark Matter}",
    eprint = "2412.09908",
    archivePrefix = "arXiv",
    journal = {Phys. Rev. Lett.},
    volume = {135},
  issue = {6},
  pages = {061002},
  numpages = {7},
  year = {2025},
  month = {Aug},
  doi = {10.1103/9dqj-q6mt},
  url = {https://link.aps.org/doi/10.1103/9dqj-q6mt}
}

@article{Mocz:2017wlg,
    author = "Mocz, Philip and Vogelsberger, Mark and Robles, Victor
        H. and others",
    title = "{Galaxy formation with BECDM {\textendash} I. Turbulence and relaxation of idealized haloes}",
    eprint = "1705.05845",
    archivePrefix = "arXiv",
    doi = "10.1093/mnras/stx1887",
    journal = "Mon. Not. Roy. Astron. Soc.",
    volume = "471",
    number = "4",
    pages = "4559--4570",
    year = "2017"
}

@article{Levkov:2018kau,
    author = "Levkov, D. G. and Panin, A. G. and Tkachev, I. I.",
    title = "{Gravitational Bose-Einstein condensation in the kinetic regime}",
    eprint = "1804.05857",
    archivePrefix = "arXiv",
    reportNumber = "INR-TH-2018-005",
    doi = "10.1103/PhysRevLett.121.151301",
    journal = "Phys. Rev. Lett.",
    volume = "121",
    number = "15",
    pages = "151301",
    year = "2018"
}

@article{Veltmaat:2018dfz,
    author = "Veltmaat, Jan and Niemeyer, Jens C. and Schwabe, Bodo",
    title = "{Formation and structure of ultralight bosonic dark matter halos}",
    eprint = "1804.09647",
    archivePrefix = "arXiv",
    doi = "10.1103/PhysRevD.98.043509",
    journal = "Phys. Rev. D",
    volume = 98,
    number = 4,
    pages = 043509,
    year = 2018
}

@article{Schwabe:2020eac,
   author = "Schwabe, Bodo and Gosenca, Mateja and Behrens, Christoph and
                     Niemeyer, Jens C. and Easther, Richard",
    title = "{Simulating mixed fuzzy and cold dark matter}",
    eprint = "2007.08256",
    archivePrefix = "arXiv",
    doi = "10.1103/PhysRevD.102.083518",
    journal = "Phys. Rev. D",
    volume = "102",
    number = "8",
    pages = "083518",
    year = "2020"
}

@article{Schwabe:2021jne,
    author = "Schwabe, Bodo and Niemeyer, Jens C.",
    title = "{Deep Zoom-In Simulation of a Fuzzy Dark Matter Galactic Halo}",
    eprint = "2110.09145",
    archivePrefix = "arXiv",
    doi = "10.1103/PhysRevLett.128.181301",
    journal = "Phys. Rev. Lett.",
    volume = "128",
    number = "18",
    pages = "181301",
    year = "2022"
}

@article{Eggemeier:2019jsu,
    author = "Eggemeier, Benedikt and Niemeyer, Jens C.",
    title = "{Formation and mass growth of axion stars in axion miniclusters}",
    eprint = "1906.01348",
    archivePrefix = "arXiv",
    doi = "10.1103/PhysRevD.100.063528",
    journal = "Phys. Rev. D",
    volume = "100",
    number = "6",
    pages = "063528",
    year = "2019"
}

@article{Mina:2020eik,
    author = "Mina, Mattia and Mota, David F. and Winther, Hans A.",
    title = "{Solitons in the dark: First approach to non-linear structure formation with fuzzy dark matter}",
    eprint = "2007.04119",
    archivePrefix = "arXiv",
    doi = "10.1051/0004-6361/202038876",
    journal = "Astron. Astrophys.",
    volume = "662",
    pages = "A29",
    year = "2022"
}

@article{RosenbergPDG,
    author = "Rosenberg, L. J. and Rybka, G. and Safdi, B.",
    title = "{Axions and Other Similar Particles}",
    booktitle = "in Review of Particle Physics",
    url="https://pdg.lbl.gov/2025/reviews/rpp2024-rev-axions.pdf",
    journal = "Phys. Rev. D",
    volume = "110",
    pages = "030001",
    year = "2024"
}

@article{Sikivie:2006ni,
    author = "Sikivie, Pierre",
    editor = "Kuster, Markus and Raffelt, Georg and Beltran, Berta",
    title = "{Axion Cosmology}",
    eprint = "astro-ph/0610440",
    archivePrefix = "arXiv",
    reportNumber = "UFIFT-HEP-06-16",
    doi = "10.1007/978-3-540-73518-2_2",
    journal = "Lect. Notes Phys.",
    volume = "741",
    pages = "19--50",
    year = "2008"
}

@article{Chadha-Day:2021szb,
    author = "Chadha-Day, Francesca and Ellis, John and Marsh, David J. E.",
    title = "{Axion dark matter: What is it and why now?}",
    eprint = "2105.01406",
    archivePrefix = "arXiv",
    reportNumber = "KCL-PH-TH/2021-20, CERN-TH-2021-045, IPPP/20/91",
    doi = "10.1126/sciadv.abj3618",
    journal = "Sci. Adv.",
    volume = "8",
    number = "8",
    pages = "abj3618",
    year = "2022"
}

@article{Kolb:1993zz,
    author = "Kolb, Edward W. and Tkachev, Igor I.",
    title = "{Axion miniclusters and Bose stars}",
    eprint = "hep-ph/9303313",
    archivePrefix = "arXiv",
    reportNumber = "FERMILAB-PUB-93-066-A",
    doi = "10.1103/PhysRevLett.71.3051",
    journal = "Phys. Rev. Lett.",
    volume = "71",
    pages = "3051--3054",
    year = "1993"
}

@article{Kolb:1993hw,
    author = "Kolb, Edward W. and Tkachev, Igor I.",
    title = "{Nonlinear axion dynamics and formation of cosmological pseudosolitons}",
    eprint = "astro-ph/9311037",
    archivePrefix = "arXiv",
    reportNumber = "FERMILAB-PUB-93-335-A",
    doi = "10.1103/PhysRevD.49.5040",
    journal = "Phys. Rev. D",
    volume = "49",
    pages = "5040--5051",
    year = "1994"
}

@article{Vaquero:2018tib,
    author = "Vaquero, Alejandro and Redondo, Javier and Stadler, Julia",
    title = "{Early seeds of axion miniclusters}",
    eprint = "1809.09241",
    archivePrefix = "arXiv",
    doi = "10.1088/1475-7516/2019/04/012",
    journal = "JCAP",
    volume = "1904",
    pages = "012",
    year = "2019"
}

@article{Buschmann:2019icd,
    author = "Buschmann, Malte and Foster, Joshua W. and Safdi, Benjamin R.",
    title = "{Early-Universe Simulations of the Cosmological Axion}",
    eprint = "1906.00967",
    archivePrefix = "arXiv",
     reportNumber = "LCTP-19-08",
    doi = "10.1103/PhysRevLett.124.161103",
    journal = "Phys. Rev. Lett.",
    volume = "124",
    number = "16",
    pages = "161103",
    year = "2020"
}

@article{Ellis:2020gtq,
    author = "Ellis, David and Marsh, David J. E. and Behrens, Christoph",
    title = "{Axion Miniclusters Made Easy}",
    eprint = "2006.08637",
    archivePrefix = "arXiv",
    doi = "10.1103/PhysRevD.103.083525",
    journal = "Phys. Rev. D",
    volume = "103",
    number = "8",
    pages = "083525",
    year = "2021"
}

@unpublished{Pierobon:2023ozb,
   author = "Pierobon, Giovanni and Redondo, Javier and Saikawa,
                   Ken'ichi and Vaquero, Alejandro and Moore, Guy D.", 
    title = "{Miniclusters from axion string simulations}",
    eprint = "2307.09941",
    archivePrefix = "arXiv",
    month = "7",
}

@article{Marsh:2015xka,
    author = "Marsh, David J. E.",
    title = "{Axion Cosmology}",
    eprint = "1510.07633",
    archivePrefix = "arXiv",
    reportNumber = "KCL-PH-TH-2015-50",
    doi = "10.1016/j.physrep.2016.06.005",
    journal = "Phys. Rept.",
    volume = "643",
    pages = "1--79",
    year = "2016"
}

@article{Ferreira:2020fam,
    author = "Ferreira, Elisa G. M.",
    title = "{Ultra-light dark matter}",
    eprint = "2005.03254",
    archivePrefix = "arXiv",
    doi = "10.1007/s00159-021-00135-6",
    journal = "Astron. Astrophys. Rev.",
    volume = "29",
    number = "1",
    pages = "7",
    year = "2021"
}

@unpublished{Eberhardt:2025caq,
    author = "Eberhardt, Andrew and Ferreira, Elisa G. M.",
    title = "{Ultralight fuzzy dark matter review}",
    eprint = "2507.00705",
    archivePrefix = "arXiv",
    month = "7",
}

@article{Hui:2021tkt,
    author = "Hui, Lam",
    title = "{Wave Dark Matter}",
    eprint = "2101.11735",
    archivePrefix = "arXiv",
    doi = "10.1146/annurev-astro-120920-010024",
    journal = "Ann. Rev. Astron. Astrophys.",
    volume = "59",
    pages = "247--289",
    year = "2021"
}

@article{Sikivie:2009qn,
    author = "Sikivie, P. and Yang, Q.",
    title = "{Bose-Einstein Condensation of Dark Matter Axions}",
    eprint = "0901.1106",
    archivePrefix = "arXiv",
    reportNumber = "UFIFT-HEP-09-1",
    doi = "10.1103/PhysRevLett.103.111301",
    journal = "Phys. Rev. Lett.",
    volume = "103",
    pages = "111301",
    year = "2009"
}

@article{Erken:2011dz,
    author = "Erken, O. and Sikivie, P. and Tam, H. and Yang, Q.",
    title = "{Cosmic axion thermalization}",
    eprint = "1111.1157",
    archivePrefix = "arXiv",
    doi = "10.1103/PhysRevD.85.063520",
    journal = "Phys. Rev. D",
    volume = 85,
    pages = 063520,
    year = 2012
}

@article{Preskill:1982cy,
    author = "Preskill, John and Wise, Mark B. and Wilczek, Frank",
    editor = "Srednicki, M. A.",
    title = "{Cosmology of the Invisible Axion}",
    reportNumber = "HUTP-82-A048, NSF-ITP-82-103",
    doi = "10.1016/0370-2693(83)90637-8",
    journal = "Phys. Lett. B",
    volume = "120",
    pages = "127--132",
    year = "1983"
}

@article{Abbott:1982af,
    author = "Abbott, L. F. and Sikivie, P.",
    editor = "Srednicki, M. A.",
    title = "{A Cosmological Bound on the Invisible Axion}",
    reportNumber = "PRINT-82-0695 (BRANDEIS)",
    doi = "10.1016/0370-2693(83)90638-X",
    journal = "Phys. Lett. B",
    volume = "120",
    pages = "133--136",
    year = "1983"
}

@article{Chen:2020cef,
    author = "Chen, Jiajun and Du, Xiaolong and Lentz, Erik W. and Marsh,
                      David J. E. and Niemeyer, Jens C.", 
    title = "{New insights into the formation and growth of boson stars in dark matter halos}",
    eprint = "2011.01333",
    archivePrefix = "arXiv",
    doi = "10.1103/PhysRevD.104.083022",
    journal = "Phys. Rev. D",
    volume = 104,
    number = 8,
    pages = 083022,
    year = 2021
}

@article{Chen:2021oot,
    author = "Chen, Jiajun and Du, Xiaolong and Lentz, Erik W. and Marsh, David J. E.",
    title = "{Relaxation times for Bose-Einstein condensation by self-interaction and gravity}",
    eprint = "2109.11474",
    archivePrefix = "arXiv",
    doi = "10.1103/PhysRevD.106.023009",
    journal = "Phys. Rev. D",
    volume = "106",
    number = "2",
    pages = "023009",
    year = "2022"
}

@article{Chan:2022bkz,
    author = "Chan, James Hung-Hsu and Sibiryakov, Sergey and Xue, Wei",
    title = "{Condensation and evaporation of boson stars}",
    eprint = "2207.04057",
    archivePrefix = "arXiv",
    doi = "10.1007/JHEP01(2024)071",
    journal = "JHEP",
    volume = "2401",
    pages = "071",
    year = "2024"
}

@article{Dmitriev:2023ipv,
    author = "Dmitriev, A. S. and Levkov, D. G. and Panin, A. G. and Tkachev, I. I.",
    title = "{Self-Similar Growth of Bose Stars}",
    eprint = "2305.01005",
    archivePrefix = "arXiv",
    reportNumber = "INR-TH-2023-006",
    doi = "10.1103/PhysRevLett.132.091001",
    journal = "Phys. Rev. Lett.",
    volume = "132",
    number = "9",
    pages = "091001",
    year = "2024"
}

@article{Svistunov:1991,
     author         = "Svistunov, B. V.",
     title          = "{Highly nonequilibrium Bose condensation in a
                  weakly interacting gas}",
     journal        = "J. Moscow Phys. Soc.",
     volume         = "1",
     year           = "1991",
     pages          = "373",
}

@article{Svistunov:2001,
title = {Strongly non-equilibrium Bose–Einstein condensation in a trapped gas},
journal = {Phys.\ Lett.\ A},
volume = {287},
eprint = "cond-mat/0009295",
archivePrefix = "arXiv",
number = {1},
pages = {169},
year = {2001},
issn = {0375-9601},
doi = {https://doi.org/10.1016/S0375-9601(01)00439-X},
url = {https://www.sciencedirect.com/science/article/pii/S037596010100439X},
author = {Boris Svistunov}
}

@article{Falkovich:1991,
title = {Nonstationary wave turbulence},
journal = {J.\ Nonlin.\ Sci.},
volume = 1,
number = 4,
pages = 457,
year = 1991,
doi = {10.1007/BF02429849},
url = {https://doi.org/10.1007/BF02429849},
author = {Falkovich, G. E. and Shafarenko, A. V.},
}

@article{Semikoz:1994zp,
      author         = "Semikoz, D. V. and Tkachev, I. I.",
      title          = "{Kinetics of Bose condensation}",
      journal        = "Phys. Rev. Lett.",
      volume         = "74",
      year           = "1995",
      pages          = "3093-3097",
      doi            = "10.1103/PhysRevLett.74.3093",
      eprint         = "hep-ph/9409202",
      archivePrefix  = "arXiv",
      reportNumber   = "FERMILAB-PUB-94-295-A",
}

@article{Semikoz:1995rd,
      author         = "Semikoz, D. V. and Tkachev, I. I.",
      title          = "{Condensation of bosons in kinetic regime}",
      journal        = "Phys. Rev.",
      volume         = "D55",
      year           = "1997",
      pages          = "489-502",
      doi            = "10.1103/PhysRevD.55.489",
      eprint         = "hep-ph/9507306",
      archivePrefix  = "arXiv",
      reportNumber   = "FERMILAB-PUB-95-220-A",
}

@article{Berges:2012us,
    author = "Berges, J. and Sexty, D.",
    title = "{Bose condensation far from equilibrium}",
    eprint = "1201.0687",
    archivePrefix = "arXiv",
    doi = "10.1103/PhysRevLett.108.161601",
    journal = "Phys. Rev. Lett.",
    volume = "108",
    pages = "161601",
    year = "2012"
}

@article{Berges:2013lsa,
    author = "Berges, J. and Boguslavski, K. and Schlichting, S. and Venugopalan, R.",
    title = "{Basin of attraction for turbulent thermalization and the range of validity of classical-statistical simulations}",
    eprint = "1312.5216",
    archivePrefix = "arXiv",
    doi = "10.1007/JHEP05(2014)054",
    journal = "JHEP",
    number="",
    volume = "1405",
    pages = "054",
    year = "2014"
}

@article{SEMISALOV2021105903,
title = {Numerical analysis of a self-similar turbulent flow in Bose–Einstein condensates},
journal = {Communications in Nonlinear Science and Numerical Simulation},
volume = {102},
pages = {105903},
year = {2021},
doi = {https://doi.org/10.1016/j.cnsns.2021.105903},
url = {https://www.sciencedirect.com/science/article/pii/S100757042100215X},
eprint={2104.14591},
archivePrefix = "arXiv",
author = {Semisalov, B.V. and Grebenev, V.N. and Medvedev, S.B. and
                  Nazarenko, S.V.},
}

@article{Kagan:1994,
     author         = "Kagan, Yu. and Svistunov, B. V.",
     title          = "{Kinetics on the onset of long-range order
                  during Bose condensation in an interacting gas}",
     journal        = "JETP",
     volume         = "78",
     year           = "1994",
     pages          = "187",
     url="http://jetp.ras.ru/cgi-bin/dn/e_078_02_0187.pdf",
}

@article{Berloff:2002,
  title = {Scenario of strongly nonequilibrated Bose-Einstein condensation},
  author = {Berloff, Natalia G. and Svistunov, Boris V.},
  journal = {Phys. Rev. A},
  volume = {66},
  issue = {1},
  pages = {013603},
  numpages = {7},
  year = {2002},
  month = {Jul},
  publisher = {American Physical Society},
  doi = {10.1103/PhysRevA.66.013603},
  url = {https://link.aps.org/doi/10.1103/PhysRevA.66.013603}
}

@article{Nowak:2011sk,
    author = "Nowak, Boris and Schole, Jan and Sexty, Denes and Gasenzer, Thomas",
    title = "{Nonthermal fixed points, vortex statistics, and superfluid turbulence in an ultracold Bose gas}",
    eprint = "1111.6127",
    archivePrefix = "arXiv",
    doi = "10.1103/PhysRevA.85.043627",
    journal = "Phys. Rev. A",
    volume = "85",
    pages = "043627",
    year = "2012"
}

@article{Nowak:2012gd,
    author = "Nowak, Boris and Gasenzer, Thomas",
    title = "{Universal dynamics on the way to thermalization}",
    eprint = "1206.3181",
    archivePrefix = "arXiv",
     doi = "10.1088/1367-2630/16/9/093052",
    journal = "New J. Phys.",
    volume = "16",
    number = "9",
    pages = "093052",
    year = "2014"
}

@article{Chantesana:2018qsb,
    author = "Chantesana, Isara and Pi\~neiro Orioli, Asier and Gasenzer, Thomas",
    title = "{Kinetic theory of nonthermal fixed points in a Bose gas}",
    eprint = "1801.09490",
    archivePrefix = "arXiv",
    doi = "10.1103/PhysRevA.99.043620",
    journal = "Phys. Rev. A",
    volume = "99",
    number = "4",
    pages = "043620",
    year = "2019"
}

@article{Mikheev:2018adp,
    author = "Mikheev, Aleksandr N. and Schmied, Christian-Marcel and Gasenzer, Thomas",
    title = "{Low-energy effective theory of nonthermal fixed points in a multicomponent Bose gas}",
    eprint = "1807.10228",
    archivePrefix = "arXiv",
    doi = "10.1103/PhysRevA.99.063622",
    journal = "Phys. Rev. A",
    volume = "99",
    number = "6",
    pages = "063622",
    year = "2019"
}

@article{Schmied:2018mte,
    author = "Schmied, Christian-Marcel and Mikheev, Aleksandr N. and Gasenzer, Thomas",
    title = "{Non-thermal fixed points: Universal dynamics far from equilibrium}",
    eprint = "1810.08143",
    archivePrefix = "arXiv",
    doi = "10.1142/S0217751X19410069",
    journal = "Int. J. Mod. Phys. A",
    volume = "34",
    number = "29",
    pages = "1941006",
    year = "2019"
}

@article{Kurkela:2012hp,
    author = "Kurkela, Aleksi and Moore, Guy D.",
    title = "{UV Cascade in Classical Yang-Mills Theory}",
    eprint = "1207.1663",
    archivePrefix = "arXiv",
    reportNumber = "INT-PUB-12-032",
    doi = "10.1103/PhysRevD.86.056008",
    journal = "Phys. Rev. D",
    volume = "86",
    pages = "056008",
    year = "2012"
}

@article{Berges:2008wm,
    author = "Berges, Juergen and Rothkopf, Alexander. and Schmidt, Jonas",
    title = "{Non-thermal fixed points: Effective weak-coupling for strongly correlated systems far from equilibrium}",
    eprint = "0803.0131",
    archivePrefix = "arXiv",
    reportNumber = "TKYNT-08-02, DESY-08-020",
    doi = "10.1103/PhysRevLett.101.041603",
    journal = "Phys. Rev. Lett.",
    volume = "101",
    pages = "041603",
    year = "2008"
}

@article{Berges:2013eia,
    author = "Berges, J. and Boguslavski, K. and Schlichting, S. and Venugopalan, R.",
    title = "{Turbulent thermalization process in heavy-ion collisions at ultrarelativistic energies}",
    eprint = "1303.5650",
    archivePrefix = "arXiv",
    doi = "10.1103/PhysRevD.89.074011",
    journal = "Phys. Rev. D",
    volume = "89",
    number = "7",
    pages = "074011",
    year = "2014"
}

@article{Berges:2013fga,
    author = "Berges, Juergen and Boguslavski, Kirill and Schlichting, Soeren and Venugopalan, Raju",
    title = "{Universal attractor in a highly occupied non-Abelian plasma}",
    eprint = "1311.3005",
    archivePrefix = "arXiv",
    doi = "10.1103/PhysRevD.89.114007",
    journal = "Phys. Rev. D",
    volume = "89",
    number = "11",
    pages = "114007",
    year = "2014"
}

@article{Berges:2014bba,
    author = "Berges, J. and Boguslavski, K. and Schlichting, S. and Venugopalan, R.",
    title = "{Universality far from equilibrium: From superfluid Bose
                  gases to heavy-ion collisions}",
    eprint = "1408.1670",
    archivePrefix = "arXiv",
    doi = "10.1103/PhysRevLett.114.061601",
    journal = "Phys. Rev. Lett.",
    volume = "114",
    number = "6",
    pages = "061601",
    year = "2015"
}

@article{AbraaoYork:2014hbk,
    author = "Abraao York, Mark C. and Kurkela, Aleksi and Lu, Egang and Moore, Guy D.",
    title = "{UV cascade in classical Yang-Mills theory via kinetic theory}",
    eprint = "1401.3751",
    archivePrefix = "arXiv",
    reportNumber = "CERN-PH-TH-2014-006",
    doi = "10.1103/PhysRevD.89.074036",
    journal = "Phys. Rev. D",
    volume = "89",
    number = "7",
    pages = "074036",
    year = "2014"
}

@article{PineiroOrioli:2015cpb,
    author = "Pi\~neiro Orioli, A. and Boguslavski, K. and Berges, J.",
    title = "{Universal self-similar dynamics of relativistic and nonrelativistic field theories near nonthermal fixed points}",
    eprint = "1503.02498",
    archivePrefix = "arXiv",
    doi = "10.1103/PhysRevD.92.025041",
    journal = "Phys. Rev. D",
    volume = "92",
    number = "2",
    pages = "025041",
    year = "2015"
}

@article{Kurkela:2015qoa,
    author = "Kurkela, Aleksi and Zhu, Yan",
    title = "{Isotropization and hydrodynamization in weakly coupled heavy-ion collisions}",
    eprint = "1506.06647",
    archivePrefix = "arXiv",
    reportNumber = "CERN-PH-TH-2015-142",
    doi = "10.1103/PhysRevLett.115.182301",
    journal = "Phys. Rev. Lett.",
    volume = "115",
    number = "18",
    pages = "182301",
    year = "2015"
}

@article{Prufer:2018hto,
    author = {Pr\"ufer, Maximilian and Kunkel, Philipp and Strobel,
                  Helmut and others},
    title = "{Observation of universal dynamics in a spinor Bose gas far from equilibrium}",
    eprint = "1805.11881",
    archivePrefix = "arXiv",
    doi = "10.1038/s41586-018-0659-0",
    journal = "Nature",
    volume = "563",
    number = "7730",
    pages = "217--220",
    year = "2018"
}

@article{Erne:2018gmz,
    author = {Erne, Sebastian and B\"ucker, Robert and Gasenzer,
                  Thomas and Berges, J\"urgen and Schmiedmayer,
                  J\"org},
    title = "{Universal dynamics in an isolated one-dimensional Bose gas far from equilibrium}",
    eprint = "1805.12310",
    archivePrefix = "arXiv",
    doi = "10.1038/s41586-018-0667-0",
    journal = "Nature",
    volume = "563",
    number = "7730",
    pages = "225--229",
    year = "2018"
}

@article{Glidden:2020qmu,
    author = "Glidden, Jake A. P. and Eigen, Christoph and Dogra, Lena
                  H. and others",
    title = "{Bidirectional dynamic scaling in an isolated Bose gas far from equilibrium}",
    eprint = "2006.01118",
    archivePrefix = "arXiv",
    doi = "10.1038/s41567-020-01114-x",
    journal = "Nature Phys.",
    volume = "17",
    number = "4",
    pages = "457--461",
    year = "2021"
}

@article{GarciaOrozco:2021,
  title = {Universal dynamics of a turbulent superfluid Bose gas},
  author = {Garc\'{\i}a-Orozco, A. D. and Madeira, L. and
                  Moreno-Armijos, M. A. and others},
  journal = {Phys. Rev. A},
  volume = {106},
  eprint = "2107.07421",
  archivePrefix = "arXiv",  
  issue = {2},
  pages = {023314},
  numpages = {10},
  year = {2022},
  month = {Aug},
  publisher = {American Physical Society},
  doi = {10.1103/PhysRevA.106.023314},
  url = {https://link.aps.org/doi/10.1103/PhysRevA.106.023314}
}

@Article{Madeira:2022,
author = {Madeira, Lucas and Bagnato, Vanderlei S.},
title = {Non-Thermal Fixed Points in Bose Gas Experiments},
journal = {Symmetry},
volume = {14},
year = {2022},
number = {4},
pages = {678},
eprint = "2203.14752",
archivePrefix = "arXiv",  
url = {https://www.mdpi.com/2073-8994/14/4/678},
issn = {2073-8994},
doi = {10.3390/sym14040678}
}

@book{Nazarenko:2011,
  title={Wave turbulence},
  author={Nazarenko, Sergey},
  series={Lecture Notes in Physics},
  volume={825},
  doi={10.1007/978-3-642-15942-8},
  isbn={978-3-642-15942-8},
  year={2011},
  publisher={Springer Berlin, Heidelberg}
}

@book{zakharov2012kolmogorov,
  title={Kolmogorov Spectra of Turbulence I: Wave Turbulence},
  author={Zakharov, V.E. and L'vov, V.S. and Falkovich, G.},
  isbn={978-3-642-50054-1},
  doi={https://doi.org/10.1007/978-3-642-50052-7},
  year={2012},
  publisher={Springer Berlin, Heidelberg}
}

@article{Zakharov:1966,
title = {Weak-turbulence spectrum in a plasma without a magnetic field},
journal = {JETP},
volume = {24},
number = {2},
pages = {455},
year = {1966},
url = {http://jetp.ras.ru/cgi-bin/dn/e_024_02_0455.pdf},
author = {Zakharov, V. E.},
}

@ARTICLE{Semisalov:2021,
       author = {{Semisalov}, B.~V. and {Grebenev}, V.~N. and {Medvedev}, S.~B. and {Nazarenko}, S.~V.},
        title = "{Numerical analysis of a self-similar turbulent flow in Bose-Einstein condensates}",
      journal = {Communications in Nonlinear Science and Numerical Simulations},
         year = 2021,
        month = nov,
       volume = {102},
          eid = {105903},
        pages = {105903},
          doi = {10.1016/j.cnsns.2021.105903},
archivePrefix = {arXiv},
       eprint = {2104.14591},
       adsurl = {https://ui.adsabs.harvard.edu/abs/2021CNSNS.10205903S}
}

@article{Schmied:2018upn,
    author = "Schmied, Christian-Marcel and Mikheev, Aleksandr N. and Gasenzer, Thomas",
    title = "{Prescaling in a far-from-equilibrium Bose gas}",
    eprint = "1807.07514",
    archivePrefix = "arXiv",
    doi = "10.1103/PhysRevLett.122.170404",
    journal = "Phys. Rev. Lett.",
    volume = "122",
    number = "17",
    pages = "170404",
    year = "2019"
}

@article{Heller:2023mah,
    author = "Heller, Michal P. and Mazeliauskas, Aleksas and Preis, Thimo",
    title = "{Prescaling Relaxation to Nonthermal Attractors}",
    eprint = "2307.07545",
    archivePrefix = "arXiv",
    doi = "10.1103/PhysRevLett.132.071602",
    journal = "Phys. Rev. Lett.",
    volume = "132",
    number = "7",
    pages = "071602",
    year = "2024"
}

@article{Galtier:2021ovg,
    author = "Galtier, Sebastien and Nazarenko, Sergey V.",
    title = "{Direct Evidence of a Dual Cascade in Gravitational Wave Turbulence}",
    eprint = "2108.09158",
    archivePrefix = "arXiv",
     doi = "10.1103/PhysRevLett.127.131101",
    journal = "Phys. Rev. Lett.",
    volume = "127",
    number = "13",
    pages = "131101",
    year = "2021"
}

@book{pitaevskii2012physical,
  title={Course of Theoretical Physics, Vol. 10: Physical Kinetics},
  author={Lifshitz, E.M. and Pitaevskii, L.P.},
  isbn={978-0-08-026480-6},
  lccn={80042162},
  url={https://www.elsevier.com/books/physical-kinetics/pitaevskii/978-0-08-026480-6},
  year={2012},
  publisher={Elsevier Science}
}

@ARTICLE{2013PhyU...56...49Z,
       author = {{Zakharov}, Vladimir E. and {Karas'}, Vyacheslav I.},
        title = "{Nonequilibrium Kolmogorov-type particle distributions and their applications}",
      journal = {Physics Uspekhi},
         year = 2013,
        month = jan,
       volume = {56},
       number = {1},
        pages = {49},
          doi = {10.3367/UFNe.0183.201301c.0055},
       adsurl = {https://ui.adsabs.harvard.edu/abs/2013PhyU...56...49Z}
}

@unpublished{Blum:2025aaa,
    author = "Blum, Kfir and Gorghetto, Marco and Hardy, Edward and Teodori, Luca",
    title = "{Bracketing the soliton-halo relation of ultralight dark matter}",
    eprint = "2504.16202",
    archivePrefix = "arXiv",
    month = "4",
}

@article{Chavanis:2011zi,
    author = "Chavanis, Pierre-Henri",
    title = "{Mass-radius relation of Newtonian self-gravitating Bose-Einstein condensates with short-range interactions: I. Analytical results}",
    eprint = "1103.2050",
    archivePrefix = "arXiv",
    doi = "10.1103/PhysRevD.84.043531",
    journal = "Phys. Rev. D",
    volume = "84",
    pages = "043531",
    year = "2011"
}

@article{Eby:2015hsq,
    author = "Eby, Joshua and Kouvaris, Chris and Nielsen, Niklas Gr{\o}nlund and Wijewardhana, L. C. R.",
    title = "{Boson Stars from Self-Interacting Dark Matter}",
    eprint = "1511.04474",
    archivePrefix = "arXiv",
    doi = "10.1007/JHEP02(2016)028",
    journal = "JHEP",
    volume = "1602",
    pages = "028",
    year = "2016",
}

@article{Levkov:2016rkk,
      author         = "Levkov, D. G. and Panin, A. G. and Tkachev, I. I.",
      title          = "{Relativistic axions from collapsing Bose stars}",
      journal        = "Phys. Rev. Lett.",
      volume         = "118",
      year           = "2017",
      number         = "1",
      pages          = "011301",
      doi            = "10.1103/PhysRevLett.118.011301",
      eprint         = "1609.03611",
      archivePrefix  = "arXiv",
      reportNumber   = "INR-TH-2016-034",
}

@article{Eby:2019ntd,
    author = "Eby, Joshua and Leembruggen, Madelyn and Street, Lauren and Suranyi, Peter and Wijewardhana, L. C. R.",
    title = "{Global view of QCD axion stars}",
    eprint = "1905.00981",
    archivePrefix = "arXiv",
    doi = "10.1103/PhysRevD.100.063002",
    journal = "Phys. Rev. D",
    volume = "100",
    number = "6",
    pages = "063002",
    year = "2019"
}

@article{Dmitriev:2021utv,
   author = "Dmitriev, A. S. and Levkov, D. G. and Panin, A. G.
       and Pushnaya, E. K. and Tkachev, I. I.",
    title = "{Instability of rotating Bose stars}",
    eprint = "2104.00962",
    archivePrefix = "arXiv",
    reportNumber = "INR-TH-2020-045",
    doi = "10.1103/PhysRevD.104.023504",
    journal = "Phys. Rev. D",
    volume = "104",
    number = "2",
    pages = "023504",
    year = "2021"
}

@article{Visinelli:2021uve,
    author = "Visinelli, Luca",
    title = "{Boson stars and oscillatons: A review}",
    eprint = "2109.05481",
    archivePrefix = "arXiv",
     doi = "10.1142/S0218271821300068",
    journal = "Int. J. Mod. Phys. D",
    volume = "30",
    number = "15",
    pages = "2130006",
    year = "2021"
}

@article{Chan:2023crj,
    author = "Chan, James Hung-Hsu and Sibiryakov, Sergey and Xue, Wei",
    title = "{Boson star normal modes}",
    eprint = "2304.13054",
    archivePrefix = "arXiv",
    doi = "10.1007/JHEP08(2023)045",
    journal = "JHEP",
    volume = "2308",
    pages = "045",
    year = "2023"
}

@article{Salasnich:2025nrt,
    author = "Salasnich, Luca and Yakimenko, Alexander",
    title = "{Collective excitations of self-gravitating ultralight dark matter cores}",
    eprint = "2501.06891",
    archivePrefix = "arXiv",
    doi = "10.1016/j.dark.2025.101973",
    journal = "Phys. Dark Univ.",
    volume = "49",
    pages = "101973",
    year = "2025"
}

@book{Rubakov:2002fi,
    author = "Rubakov, V. A.",
    title = "{Classical theory of gauge fields}",
    isbn = "978-0-691-05927-3, 978-0-691-05927-3",
    publisher = "Princeton University Press",
    address = "Princeton, New Jersey",
    month = "5",
    year = "2002"
}

@article{Maseizik:2024qly,
    author = {Maseizik, Dennis and Sigl, G{\"u}nter},
    title = "{Distributions and collision rates of ALP stars in the Milky~Way}",
    eprint = "2404.07908",
    archivePrefix = "arXiv",
    doi = "10.1103/PhysRevD.110.083015",
    journal = "Phys. Rev. D",
    volume = "110",
    number = "8",
    pages = "083015",
    year = "2024"
}

@article{Teukolsky:1999rm,
    author = "Teukolsky, Saul A.",
    title = "{On the stability of the iterated Crank-Nicholson method in numerical relativity}",
    eprint = "gr-qc/9909026",
    archivePrefix = "arXiv",
    doi = "10.1103/PhysRevD.61.087501",
    journal = "Phys. Rev. D",
    volume = "61",
    pages = "087501",
    year = "2000"
}

@book{NR,
  author = {Press, W.H. and Teukolsky, S.A. and Vetterling, W.T. and Flannery, B.P.},
  edition = 3,
  isbn = 9780521880688,
  lccn = 89015841,
  publisher = {Cambridge University Press},
  title = {Numerical Recipes: The Art of Scientific Computing},
  url_my = {https://books.google.ru/books?id=1aAOdzK3FegC&printsec=frontcover&hl=ru&source=gbs_ge_summary_r&cad=0#v=onepage&q&f=false},
  year = 2007
}

@article{Ruffini:1969qy,
      author         = "Ruffini, Remo and Bonazzola, Silvano",
      title          = "{Systems of selfgravitating particles in general
                        relativity and the concept of an equation of state}",
      journal        = "Phys. Rev.",
      volume         = "187",
      year           = "1969",
      pages          = "1767-1783",
      doi            = "10.1103/PhysRev.187.1767",
}

@article{Tkachev:1986tr,
      author         = "Tkachev, I. I.",
      title          = "{Coherent scalar field oscillations forming compact
                        astrophysical objects}",
      journal        = "Sov. Astron. Lett.",
      volume         = "12",
      year           = "1986",
      pages          = "305-308",
}

@article{Micha:2002ey,
    author = "Micha, Raphael and Tkachev, Igor I.",
    title = "{Relativistic turbulence: A Long way from preheating to equilibrium}",
    eprint = "hep-ph/0210202",
    archivePrefix = "arXiv",
    doi = "10.1103/PhysRevLett.90.121301",
    journal = "Phys. Rev. Lett.",
    volume = "90",
    pages = "121301",
    year = "2003"
}

@article{Micha:2004bv,
    author = "Micha, Raphael and Tkachev, Igor I.",
    title = "{Turbulent thermalization}",
    eprint = "hep-ph/0403101",
    archivePrefix = "arXiv",
    doi = "10.1103/PhysRevD.70.043538",
    journal = "Phys. Rev. D",
    volume = "70",
    pages = "043538",
    year = "2004"
}

@article{Nori:2020jzx,
    author = "Nori, Matteo and Baldi, Marco",
    title = "{Scaling relations of fuzzy dark matter haloes \textendash{} I. Individual systems in their cosmological environment}",
    eprint = "2007.01316",
    archivePrefix = "arXiv",
    doi = "10.1093/mnras/staa3772",
    journal = "Mon. Not. Roy. Astron. Soc.",
    volume = "501",
    number = "1",
    pages = "1539--1556",
    year = "2021"
}

@article{Chan:2021bja,
author = "Chan, Hei Yin Jowett and others",
    title = "{The diversity of core\textendash{}halo structure in the fuzzy dark matter model}",
    eprint = "2110.11882",
    archivePrefix = "arXiv",
    doi = "10.1093/mnras/stac063",
    journal = "MNRAS",
    volume = "511",
    number = "1",
    pages = "943",
    year = "2022"
}

@article{Maseizik:2024uln,
    author = {Maseizik, Dennis and Mondal, Sagnik and Seong, Hyeonseok and Sigl, G{\"u}nter},
    title = "{Radio lines from accreting axion stars}",
    eprint = "2409.13121",
    archivePrefix = "arXiv",
    reportNumber = "DESY-24-140",
    doi = "10.1088/1475-7516/2025/05/033",
    journal = "JCAP",
    volume = 2505,
    pages = 033,
    year = 2025
}

@article{Gorghetto:2024vnp,
    author = "Gorghetto, Marco and Hardy, Edward and Villadoro, Giovanni",
    title = "{More axion stars from strings}",
    eprint = "2405.19389",
    archivePrefix = "arXiv",
    reportNumber = "DESY-24-075",
    doi = "10.1007/JHEP08(2024)126",
    journal = "JHEP",
    volume = "2408",
    pages = "126",
    year = "2024"
}

@article{Dmitriev:2023jnu,
    author = "Dmitriev, A. S. and Dmitrieva, E. A. and Panin, A. G.",
    title = "{Scattering of Linear Waves on a Soliton}",
    eprint = "2312.05867",
    archivePrefix = "arXiv",
    reportNumber = "INR-TH-2023-022",
    doi = "10.1134/S0021364024600319",
    journal = "JETP Lett.",
    volume = "119",
    number = "5",
    pages = "395--401",
    year = "2024"
}

@article{Chavanis:2025qcg,
    author = "Chavanis, Pierre-Henri",
    title = "{A review of basic results on the Bose{\textendash}Einstein condensate dark matter model}",
    doi = "10.3389/fspas.2025.1538434",
    journal = "Front. Astron. Space Sci.",
    volume = "12",
    pages = "1538434",
    year = "2025"
}

@article{Chavanis:2020upb,
    author = "Chavanis, Pierre-Henri",
    title = "{Landau equation for self-gravitating classical and quantum particles: application to dark matter}",
    eprint = "2012.12858",
    archivePrefix = "arXiv",
    doi = "10.1140/epjp/s13360-021-01617-3",
    journal = "Eur. Phys. J. Plus",
    volume = "136",
    number = "6",
    pages = "703",
    year = "2021"
}

@article{Amin:2025sla,
    author = "Amin, Mustafa A. and May, Simon and Mirbabayi, Mehrdad",
    title = "{Early growth of structure in warm wave dark matter}",
    eprint = "2506.12131",
    archivePrefix = "arXiv",
    doi = "10.1088/1475-7516/2025/10/040",
    journal = "JCAP",
    volume = "10",
    pages = "040",
    year = "2025"
}

@article{Amin:2025nxm,
    author = "Amin, Mustafa A. and Delos, M. Sten",
    title = "{Growth of Structure in Multi-species Wave Dark Matter}",
    eprint = "2510.17977",
    archivePrefix = "arXiv",
}


\end{document}